\documentclass[twocolumn]{aastex61}
\usepackage[version=4]{mhchem}

\newcommand\omcfe{\mbox{[(O$-$C)/Fe]}}
\newcommand\mathomcfe{\mathrm{\omcfe}}
\newcommand\teff{$T_\mathrm{eff}$}
\newcommand\mteff{T_\mathrm{eff}} 
\newcommand\logg{$\log g$}
\newcommand\mlogg{\log g}
\newcommand\feh{[Fe/H]}
\newcommand\mfeh{\mathrm{[Fe/H]}}
\newcommand\mh{[M/H]}
\newcommand\afe{[$\alpha$/Fe]}
\newcommand\tife{[Ti/Fe]}
\newcommand\mtife{\mathrm{[Ti/Fe]}}
\newcommand\cfe{[C/Fe]}
\newcommand\ofe{[O/Fe]}
\newcommand\cm{[C/M]}
\newcommand\om{[O/M]}
\newcommand\am{[$\alpha$/M]}
\newcommand\al{$\alpha$}
\newcommand\chisq{$\chi^2$}
\newcommand\fei{\ion{Fe}{1}}
\newcommand\tii{\ion{Ti}{1}}

\accepted{October 25, 2017}
\submitjournal{ApJ}

\shorttitle{Effective temperature, metallicity, and Ti abundance of M dwarfs}
\shortauthors{Veyette et al.}

\begin{document}

\title{A physically motivated and empirically calibrated method to measure\\ effective temperature, metallicity, and Ti abundance of M dwarfs}

\correspondingauthor{Mark J. Veyette}
\email{mveyette@bu.edu}

\author{Mark J. Veyette}
\affiliation{Department of Astronomy \& Institute for Astrophysical Research, Boston University, 725 Commonwealth Ave., Boston, MA, 02215 USA}

\author{Philip S. Muirhead}
\affiliation{Department of Astronomy \& Institute for Astrophysical Research, Boston University, 725 Commonwealth Ave., Boston, MA, 02215 USA}

\author{Andrew W. Mann}
\affiliation{Department of Astronomy, The University of Texas at Austin, Austin, TX 78712, USA}
\altaffiliation{Hubble Fellow}

\author{John M. Brewer}
\affiliation{Department of Astronomy, Yale University, 52 Hillhouse Avenue, New Haven, CT 06511, USA}

\author{France Allard}
\affiliation{Centre de Recherche Astrophysique de Lyon, UMR 5574, Universit\'e de Lyon, ENS de Lyon, Universit\'e Lyon 1, CNRS, F-69007, Lyon, France}

\author{Derek Homeier}
\affiliation{Zentrum f{\"u}r Astronomie der Universit{\"a}t Heidelberg, Landessternwarte, K{\"o}nigstuhl 12, 69117 Heidelberg, German}

\begin{abstract}
The ability to perform detailed chemical analysis of Sun-like F-, G-, and K-type stars is a powerful tool with many applications including studying the chemical evolution of the Galaxy and constraining planet formation theories. Unfortunately, complications in modeling cooler stellar atmospheres hinders similar analysis of M-dwarf stars. Empirically-calibrated methods to measure M dwarf metallicity from moderate-resolution spectra are currently limited to measuring overall metallicity and rely on astrophysical abundance correlations in stellar populations. We present a new, empirical calibration of synthetic M dwarf spectra that can be used to infer effective temperature, Fe abundance, and Ti abundance. We obtained high-resolution (R$\sim$25,000), Y-band ($\sim$1 $\micron$) spectra of 29 M dwarfs with NIRSPEC on Keck II. Using the \texttt{PHOENIX} stellar atmosphere modeling code (version 15.5), we generated a grid of synthetic spectra covering a range of temperatures, metallicities, and alpha-enhancements. From our observed and synthetic spectra, we measured the equivalent widths of multiple \ion{Fe}{1} and \ion{Ti}{1} lines and a temperature-sensitive index based on the FeH bandhead. We used abundances measured from widely-separated solar-type companions to empirically calibrate transformations to the observed indices and equivalent widths that force agreement with the models. Our calibration achieves precisions in \teff{}, \feh{}, and \tife{} of 60 K, 0.1 dex, and 0.05 dex, respectively and is calibrated for 3200 K $<$ \teff{} $<$ 4100 K, $-0.7$ $<$ \feh{} $<$ +0.3, and $-$0.05 $<$ \tife{} $<$ +0.3. This work is a step toward detailed chemical analysis of M dwarfs at a similar precision achieved for FGK stars.
\end{abstract}

\keywords{stars: abundances --- stars: fundamental parameters --- stars: late-type --- stars: low-mass, brown dwarfs --- stars: atmospheres}

\section{Introduction}\label{intro}
Detailed spectroscopic analysis of planet-hosting stars is an important step in the follow-up characterization of exoplanetary systems. Analysis of high-resolution optical spectra of Sun-like F-, G-, and K-type stars provides accurate fundamental parameters like effective temperature, surface gravity, and chemical abundances for numerous elements. Accurate stellar parameters are necessary to characterize exoplanetary systems including the potential habitability of rocky, Earth-sized planets \citep[e.g.,][]{Everett2013}. Furthermore, detailed chemical analysis of planet-hosts allow for investigations into trends between planet-occurrence and stellar composition, which can constrain planet formation theories.

There is a well-established trend between stellar metallicity and the occurrence of giant planets around solar-type stars \citep{Gonzalez1997,Santos2001,Fischer2005}, which is consistent with the core-accretion theory of planet formation. Additionally, several studies found that stars which host giant planets are further enhanced in refractory elements like Mg, Si, and Ti over and above the observed planet-metallicity correlation \citep{Brugamyer2011,Adibekyan2012a}, further suggesting that the primordial composition of the protoplanetary disk plays a significant role in the efficiency of giant planet formation. Whether this dependence on stellar composition continues down to lower mass planets is still unclear. \citet{Wang2015} found that all planets, including Earth-sized planets ($R_p \le 1.7 R_\earth$), are more common around metal-rich stars, but that the dependence of planet occurrence on metallicity decreases for smaller planets. \citet{Adibekyan2012b} found that metal-poor stars which host Neptune-size or super-Earth planets are also overabundant in \al{}-elements compared to non-hosts. Other works, however, do not find similar trends. Based on a small sample of planets detected by radial velocity surveys, \citet{Sousa2011} did not find evidence of a planet-metallicity correlation for low-mass planets. Based on \textit{Kepler} results, \citet{Everett2013} and \citet{Buchhave2012} showed that planets with $R_p < 4 R_\earth$ exist around stars with a wide range of metallicities. \citet{Buchhave2014} claim that there does exist a moderate metallicity enhancement for stars that host planets with radii between $1.7 R_\earth$ and $4 R_\earth$, but not for terrestrial hosts. Similarly, \citet{Buchhave2015} found that the metallicity distribution of stars that host planets with $R_p \le 1.7 R_\earth$ is indistinguishable from that of non-hosts. \citet{Zhu2016} modeled the planet-metallicity correlation as a power law up to a critical metallicity and argued that the difficulty of detecting a planet-metallicity correlation for small planets is due to the combined effect of high planet occurrence rate and low detection efficiency. Their model reproduces the null detection of \citet{Buchhave2015} as well as the tentative detection of \citet{Wang2015}, suggesting that a planet-metallicity correlation for small planets cannot be ruled out.

\citet{Johnson2009} and \citet{Johnson2010} performed similar analysis on planet-hosting M dwarfs, finding that, as with Sun-like stars, Jupiter-size giant planets are more common around metal-rich M dwarfs. Unlike Sun-like stars, there is no evidence that this trend continues down to Neptune-size or smaller planets \citep{Mann2013b,Gaidos2016}. Unfortunately, due to difficulties in performing detailed chemical analysis on M dwarfs, there have been no statistical studies on the correlation between the occurrence of terrestrial planets around M dwarfs and the abundance of \textit{refractory elements}. Such studies would shed light on the role that initial composition plays in planet formation. It is increasingly important to develop new methods for detailed spectroscopic analysis of M dwarfs as many current and future planet-detection surveys specifically target M dwarfs (e.g., MEarth, \citealt{Nutzman2008}; \textit{TESS}, \citealt{Ricker2014}; HARPS surveys, e.g., \citealt{Astudillo-Defru2017}; CARMENES \citealt{Quirrenbach2010}; the Habitable Zone Planet Finder, \citealt{Mahadevan2010}; and SPIRou, \citealt{Artigau2011}).

Detailed spectroscopic analysis of M-dwarf stars is hindered by the difficulty of accurately modeling the millions of molecular lines present in M dwarf spectra, as a result of their cooler atmospheres. To avoid this issue, previous studies relied on empirical calibrations based on observations of M dwarfs with widely-separated F-, G-, or K-type binary companions \citep[e.g.,][]{Bonfils2005}. The two stars are assumed to have formed together with the same initial composition. The overall metallicity of the system  ([M/H], or [Fe/H] as proxy) can be measured from the FGK companion and used to empirically calibrate metallicity-sensitive optical-NIR colors and magnitudes \citep{Bonfils2005,Casagrande2008,Johnson2009,Schlaufman2010,Neves2012,Johnson2012,Hejazi2015,Dittman2016}, features in moderate-resolution optical or NIR spectra \citep{Rojas2010,Rojas2012,Terrien2012,Mann2013a,Newton2014}, and features in high-resolution optical spectra \citep{Pineda2013,Neves2014,Maldonado2015}.

Metallicity estimates based on empirically-calibrated features in M dwarf spectra can achieve $\sim$0.1 dex precision in [Fe/H]. However, they are not direct measurements of Fe abundance. Even those based on high-resolution spectra are not based directly on \fei{} lines. As such, these methods measure Fe abundance indirectly through astrophysical abundance correlations in stellar populations. For example, the relative abundance of C and O correlates strongly with metallicity in the solar neighborhood \citep{Delgado2010,Petigura2011,Nissen2013,Teske2014,Nissen2014,Brewer2016b}. \citet{Veyette2016b} showed that the pseudo-continuum level in M dwarfs is highly sensitive to the relative abundances of C and O. They further showed that C and O abundances are the primary mechanism behind mid-M dwarf metallicity calibrations based on moderate-resolution spectra. As indirect tracers of metallicity, empirical methods are limited by the inherent scatter in correlated abundance trends and will fail for stars with non-standard abundance ratios. 

Attempts to derive model-dependant abundances for M dwarfs have been less common. \citet{Mould1976,Mould1978} first applied the method of spectral synthesis to M dwarfs, and \citet{Valenti1998} pioneered the modern approach to derive precise M dwarf parameters through spectral synthesis at high resolution. \citet{Woolf2005} used the equivalent width (EW) matching code MOOG \citep{Sneden1973} to measure Ti and Fe abundances from atomic lines in M and K dwarfs. More recently, updated line lists and high-resolution NIR spectroscopy have allowed standard spectral analysis techniques to be applied to M dwarfs with a precision similar to such analysis of FGK stars. \citet{Tsuji2014} and \citet{Tsuji2015} measured C and O abundances of M dwarfs by comparing the equivalent widths of blended \ce{CO} and \ce{H2O} lines in high resolution K-band spectra of M dwarfs to their Unified Cloudy Models. \citet{Onehag2012}, \citet{Lindgren2016} and \citet{Lindgren2017} utilized MARCS models \citep{Gustafsson2008} and the Spectroscopy Made Easy \citep[SME][]{Valenti1996,Piskunov2017} spectral synthesis code to infer M dwarf effective temperatures and metallicities to a precision of 100 K and 0.05 dex, respectively. \citet{Souto2017} used MARCS models and the turbospectrum code \citep{Alvarez1998,Plez2012} to synthesize SDSS APOGEE spectra of two planet-hosting, early-M dwarfs (Kepler-138 and Kepler-186) and measured chemical abundances for 13 elements with a precision of order 0.1 dex. 

Current M dwarf spectral synthesis attempts, however, still suffer some drawbacks. For one, they rely on presupposing accurate stellar parameters to generate model atmospheres for spectral synthesis. Most works so far employed either empirical color-temperature relations, such as those of \citet{Casagrande2008} and \citet{Mann2015}, or empirical absolute magnitude-temperature relations. For \logg{}, many studies utilized the \logg{}-mass relation of \citet{Bean2006} and absolute magnitude-mass relations such as those of \citet{Delfosse2000} and \citet{Benedict2016}. Others calculated \logg{} using those same absolute magnitude-mass relations and radius estimates from absolute magnitude-radius relations such as those of \citet{Mann2015}. Inconsistencies in how parameters are determined for model generation could lead to inconsistencies in derived abundances.

The accuracy of abundances derived from spectral synthesis depend strongly on the accuracy of the model atmospheres used. The pervasiveness of molecular opacity and the importance of convective energy transport in cool dwarf atmospheres pose unique challenges to accurately modeling M dwarf spectra. These challenges complicate attempts to derive accurate fundamental parameters directly from spectral synthesis. Results from direct spectral synthesis are often inconsistent with result from empirical methods \citep[e.g.,][]{Passegger2016}.
Recently, \citet{2017arXiv170806211R} found that directly comparing high-resolution H-band spectra of M dwarfs to BT-Settl synthetic spectra resulted in best-fit temperatures and metallicities that differed by up to 350 K and 0.8 dex from those measured based on empirically-calibrated methods \citep{Terrien2015b}.

The overall metallicities derived by \citet{Lindgren2016} and \citet{Lindgren2017} are in good agreement with those derived from the \citet{Rojas2012}, \citet{Terrien2012}, and \citet{Mann2013a} empirical calibrations based on moderate-resolution spectra, agreeing within measurement uncertainties. Additionally, \citet{Lindgren2016} analyzed four FGK+M binaries, finding excellent agreement (0.01--0.04 dex difference) between metallicities measured independently from either component. \citet{Onehag2012}, \citet{Lindgren2016}, and \citet{Lindgren2017} did not fit for individual elemental abundances, so the accuracy of their methods for detailed chemical analysis is unknown. Furthermore, they found discrepancies between temperatures derived through their spectral synthesis and those derived from empirical calibrations. Their temperatures are consistently $\sim$100 K lower than those determined by \citet{Mann2015} which were determined by comparing optical spectra to BT-Settl models, but ultimately tuned to match long baseline optical interferometry observations.

The metallicities derived by \citet{Souto2017} of Kepler-138 and Kepler-186 are consistently $\sim$0.1--0.2 dex higher than those derived from the \citet{Rojas2012}, \citet{Terrien2012}, \citet{Mann2013a}, and \citet{Terrien2015b} empirical calibrations based on moderate-resolution spectra. The empirically-calibrated methods do not measure Fe abundance directly, but can predict M dwarf [Fe/H] to $<$ 0.1 dex precision. Analysis of more APOGEE M dwarf spectra is needed to determine if there is a statistically significant difference in metallicities determined from spectral synthesis and from empirical calibrations. No independent analysis of abundances beyond overall metallicity for Kepler-138 and Kepler-186 are available for comparison to the \citet{Souto2017} results.

Inconsistencies between empirically-calibrated and model-dependent methods for spectroscopic characterization of M dwarfs must be resolved in order to allow detailed chemical analysis of M dwarfs with a similar accuracy and precision that is achieved for FGK stars. We present here a new method to derive \teff{}, \feh{}, and \tife{} from high-resolution NIR M dwarf spectra that is both physically motivated and empirically calibrated. In Section~\ref{obs}, we describe our Keck/NIRSPEC observations of M dwarfs in FGK+M systems. In Section~\ref{abund}, we describe how our method utilizes state-of-the-art stellar atmosphere models to provide the nonlinear relations for how M dwarf spectra change as a function of stellar parameters and composition, and how our we calibrate our method with FGK+M systems. We discuss our results in Section~\ref{disc} and summarize them in Section~\ref{summary}.

\section{Observations}\label{obs}

\subsection{NIRSPEC Observations of M dwarfs}

\begin{figure*}
\centering
\includegraphics[width=\linewidth]{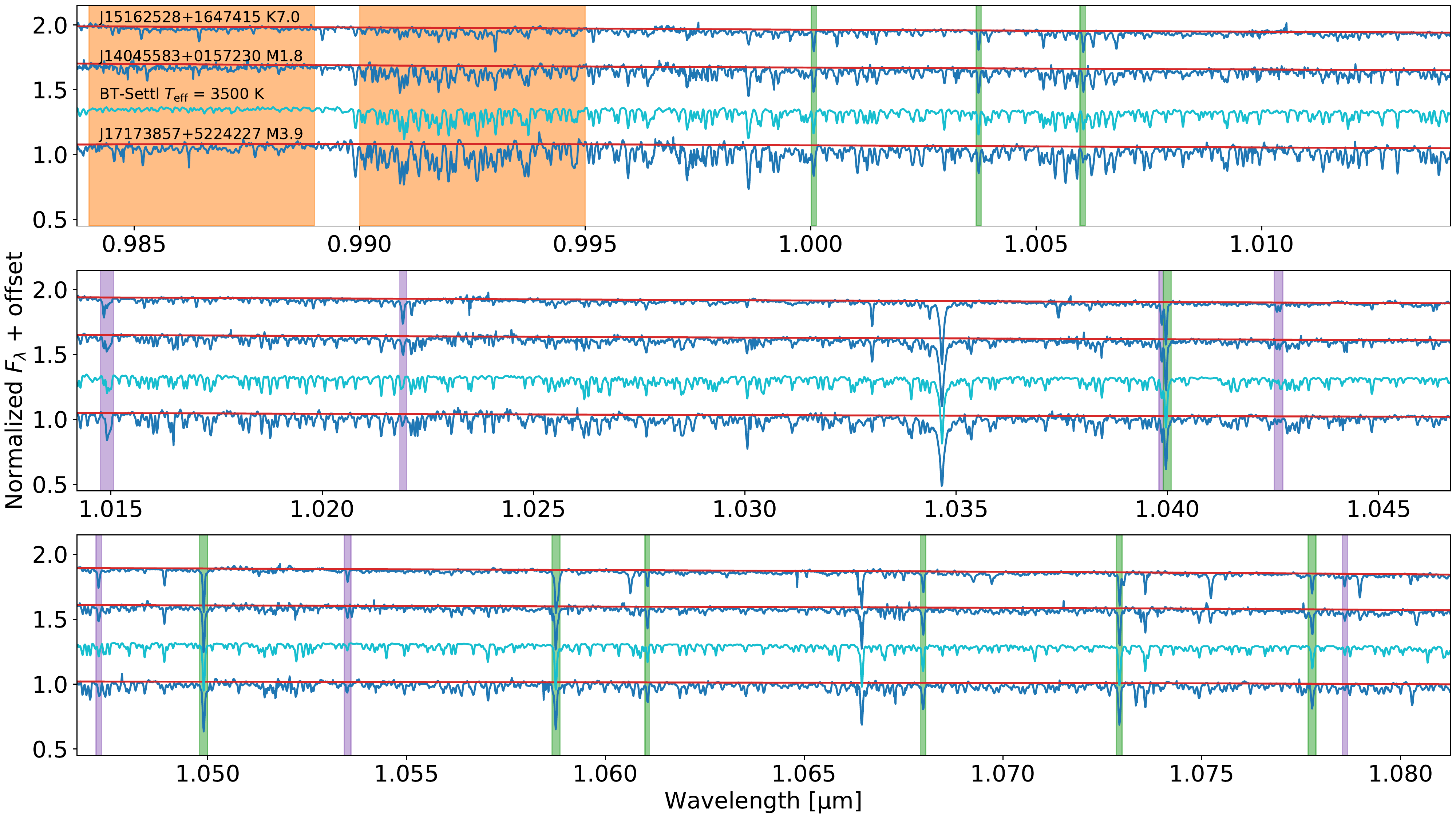}
\caption{A representative sample of our fully reduced NIRSPEC observations. The third spectrum (lighter blue) is a BT-Settl synthetic spectrum for comparison. Red lines show the pseudo-continuum level. Orange shading denotes the two regions used in the temperature-sensitive index based on the FeH bandhead. Purple shading denotes \fei{} lines used in abundance determination. Green shading denotes \tii{} lines.  \label{spectra}}
\end{figure*}

On the nights of 2016-05-24 and 2017-02-02, we used the NIRSPEC instrument \citep{McLean1998} on Keck II on Mauna Kea to observe a total of 44 M dwarfs from the \citet{Mann2013a} catalog of FGK+M systems. We observed with the NIRSPEC-1 filter covering 0.947-1.121 $\micron$, corresponding to the photometric Y band. We used the 0.432$\times$12 arcsecond slit for a spectral resolution of $R\simeq$25,000. We employed the standard ABBA slit-nodding pattern for a total of at least 8 exposures per target. We chose single-image exposure times necessary to reach a combined peak S/N $>$ 150 per pixel. We also obtained dark, flat field, and Ne-Ar-Xe-Kr arc lamp calibration images. To help remove the many telluric lines present in the NIR, an A0V star is usually observed close in time and airmass to each target. However due to the very limited number of contaminating telluric lines in Y-band, we only observed two A0V stars at two different airmasses each night. We used these observations when calibrating instrumental effects.

We used the REDSPEC\footnote{\url{https://www2.keck.hawaii.edu/inst/nirspec/redspec.html}} code to spatially and spectrally rectify each image. For initial wavelength calibration, we used sky OH lines for all orders except 72 and 74 for which we used the arc lamp lines because these orders do not contain enough OH lines. Following the procedure outlined in \citet{Cushing2004}, we optimally extracted \citep{Horne1986} the 1D spectrum from each image and combined all spectra of the same object using a variance-weighted mean. The spectra are contaminated by fringes caused by interference between the order-sorting filter and the long-wavelength blocking filter. We used Fourier filtering to remove the fringes. First, we used the Fourier transform of the A0V stars to determine the dominate frequencies of the fringes. The A0V spectra have very few stellar or telluric lines and are dominated by the fringe signal which stands out as a large peak in the frequency spectrum. We then filtered the fringe frequencies from all target spectra in Fourier space with a FIR notch filter based on a Hanning window with a width of $6\times10^{-3}$  $\mathrm{pix}^{-1}$ and centered on the peak frequency as determined from the A0V observations. This procedure is similar to an option available in the REDSPEC package to remove fringing.

Due to the fact that Y band is nearly devoid of telluric lines, we chose not to use the A0V observations for telluric correction. Instead, we corrected for the throughput of the instrument by matching our observations to publicly available\footnote{\url{https://phoenix.ens-lyon.fr/Grids/BT-Settl/AGSS2009/SPECTRA/}} BT-Settl synthetic spectra. At the same time, we used the models to improve our wavelength calibration and shift each spectrum to the rest frame. For each order of each observed spectrum, we multiplied the flux by a 3rd order Chebyshev polynomial and applied a linear correction to the wavelengths in order to best match a synthetic spectrum. The wavelength correction shifts the spectrum to the rest frame and removes extrapolation error in the REDSPEC wavelength calibration which arises due to the low number of OH and arc lamp lines in Y-band. We iterated over all models within a grid covering \teff{} = 2600--4300 K, \logg{} = 5.0, and \mh{} = $-$1.0--+0.5. For each model, we found the best fit coefficients for the throughput correction and wavelength calibration via \chisq{} minimization. We used the coefficients that produced the lowest \chisq{} over the entire model grid to apply the final flux and wavelength calibration. Figure~\ref{spectra} shows some representative samples of fully reduced spectra.

\section{Calibrating a method to measure \teff{}, \feh{}, and \tife{}}\label{abund}

We chose to combine two approaches to analyzing M dwarf spectra and developed a method that is both physically motivated and empirically calibrated. Our basic strategy is to use a grid of synthetic spectra to provide the nonlinear relations for how an M dwarf spectrum should change as a function of physical parameters, but then apply simple transformations to measured EWs and spectral indices to force agreement with observations of well-characterized FGK+M systems.

\subsection{Model grid}

We used the 2017 version of the \texttt{PHOENIX} atmosphere modeling code \citep{Allard2012a,Baraffe2015,Allard2016} to generate a grid of synthetic M dwarf spectra\footnote{All synthetic spectra are available for download online at \url{http://people.bu.edu/mveyette/phoenix/}}. Due to the many issues in modeling M dwarf spectra (see Section ~\ref{disc} for a discussion of some of these issues), we chose not to finely tune our models to recreate observed spectra and compare the model-derived parameters to those measured from empirical methods. We leave this exercise to future studies involving high-resolution spectra over a broader range of the full spectral energy distribution. Instead, we created a generalized grid of models with the goal of accurately representing the majority of main-sequence M dwarfs with the fewest number of free parameters. The most important stellar parameters for the \texttt{PHOENIX} models are the \teff{}, \logg{}, and composition.

We chose to parameterize the composition in terms of a metallicity value \mh{} that scales all elements equally from their solar abundance, and a second alpha-enhancement value \am{} that additionally scales the elements Ne, Mg, Si, S, Ar, Ca, and Ti by a single value. Unlike other model grids, we do not include O as an alpha element when varying \am{}. Instead, we treat C and O separately, parameterizing their abundance as a function of \mh{} as described below. Solar abundances are based on \citet{Asplund2009} with revisions from \citet{Caffau2011} as described in \citet{Allard2013}.

We note that $\mathrm{\feh{}} \approx \mathrm{\mh{}}$ if alpha elements are treated separately and not included when calculating the average metallicity of a star, and in our models \feh{} = \mh{} by definition. In this paper we use \mh{} when referring to the metallicity of our models and \feh{} when referring to the metallicity of individual stars as that is what has been measured. However, we consider them equivalent, assuming that \feh{} is a perfect proxy for \mh{} when alpha elements are varied independently.

\subsubsection{Treatment of C and O abundances}

\begin{figure}
\centering
\includegraphics[width=\linewidth]{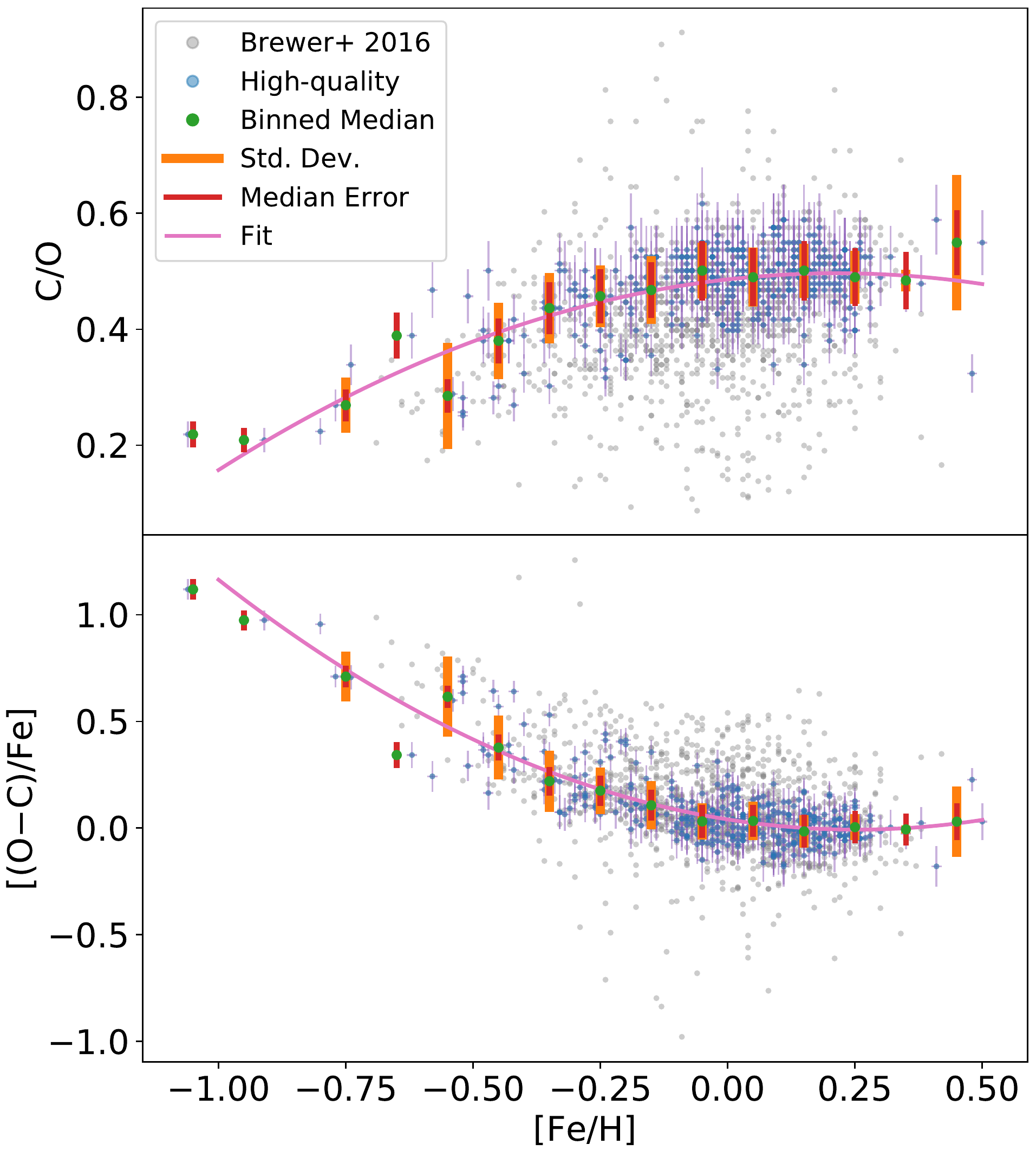}
\caption{C/O and \omcfe{} vs \feh{} for FGK stars analyzed by \citet{Brewer2016a}. High-S/N, Sun-like stars, as described in the text, are shown in blue with their measurement uncertainties. Median values, $\pm$1 standard deviation bars, and $\pm$1 median measurement error bars are shown for 0.1 dex bins in \feh{}.} Also shown are quadratic fits to the high-quality sample. There is a strong correlation between \feh{} and the relative abundance of C and O, with scatter nearly consistent with measurement errors. \label{feh_v_co}
\end{figure}

The relative abundance of C and O in an M dwarf's atmosphere has a large effect on the pseudo continuum level in its spectrum \citep{Veyette2016b}. We accounted for this effect by scaling C and O abundances as functions of metallicity when generating our model grid. Spectroscopic surveys of solar neighborhood FGK stars show a tight trend between C, O, and Fe abundances \citep{Delgado2010,Nissen2014,Teske2014,Brewer2016b}. We derived empirical relations between Fe, C, and O abundances to use when calculating our model grid based on the abundance data of \citet{Brewer2016a}, who calculated abundances of 15 elements for 1,617 FGK stars. We first rescaled the abundances to match the solar abundance scale used in the \texttt{PHOENIX} models. Figure~\ref{feh_v_co} shows how the relative abundance of C and O varies as a function of \feh{}.

In order to derive an accurate model for how C and O vary with Fe, we made several cuts to the \citet{Brewer2016a} sample. First, we limited the sample to only Sun-like stars. \citet{Brewer2016b} observed that the scatter in the measured C/O of stars in the solar neighborhood was reduced when limiting analysis to stars with \logg{} $>$ 4.0, and \teff{} within $\pm$200 K of
the Sun. They suggest this is due to larger systematic uncertainty for models of stars hotter or cooler than the Sun. \citet{Brewer2016a} fit for and removed any temperature dependence they could measure in their abundance determinations. However, they only fit to stars with \teff{} = 4800--6200 K and there still exists a noticeable temperature dependence in C/O for stars significantly hotter or cooler than the Sun. We adopted the same Sun-like criteria as \citet{Brewer2016b}. Next, we cut any stars with reported S/N $<$ 100. Finally, we add back any stars with \feh{} $<$ -0.7 or \feh{} $>$ 0.4 that also meet the S/N cut. This adds back four metal-poor late-G/early-K stars, two metal-rich late-G/early-K stars, and one metal-rich late-F/early-G star. We add these stars back because there are very few Sun-like stars at the metallicity extremes of the sample where the fit tended to C/O values that were unrealistically low compared to other surveys that focused on lower-metallicity stars \citep[e.g.][]{Nissen2014}. Although not statistically motivated, this step was necessary to ensure the fit remains realistic within the full range of our model grid. We note that the added stars lie beyond the \feh{} range of our FGK+M calibration sample ($-$0.7 $<$ \feh{} $<$ $+$0.35) and the effect on the fit is negligible over this range. This fit is not valid for \feh{} $<$ -1.

Figure~\ref{feh_v_co} shows the 341 stars that make our cuts in blue and the full sample in grey. The scatter in the \feh{}-C/O relation is significantly reduced when considering only high-S/N, Sun-like stars compared to the full sample. Also shown for the stars that make our cuts are 1-sigma error bars calculated by propagating the individual uncertainties on [C/H] and [O/H] (0.026 and 0.036 dex, respectively) from \citet{Brewer2016a}. For clarity, we also show the median C/O, the standard deviation of C/O, and the median measurement error in C/O calculated for 0.1 dex bins in \feh{}. We note that the median measurement error is not the error in the binned average, but rather represents the typical uncertainty on a single C/O measurement in a given bin. The variations in C/O as a function of \feh{} are nearly consistent with measurement errors, though there exists some evidence of inherent scatter, particularly at the low-metallicity end (see Section~\ref{disc} for more discussion).

\citet{Veyette2016b} found that the log difference in O and C abundance relative to Fe abundance and scaled from solar, \omcfe{}, is a good tracer of C and O effects on the pseudo-continuum in M dwarf spectra, with C/O being the second most important parameter. Therefore, we fit \omcfe{} and C/O as quadratic functions of \feh{}. We used an unweighted fit to the individual points in the high-quality sample, not the binned points. The resulting fits are
\begin{eqnarray}
    \mathrm{C/O} & = & 0.486 + 0.099\mathrm{\feh{}} - 0.230\mathrm{\feh{}}^2, \label{eq:ctoo}\\
    \mathomcfe{} & = & 0.040 - 0.378\mathrm{\feh{}} + 0.747\mathrm{\feh{}}^2. \label{eq:omcfe}
\end{eqnarray}
The reduced \chisq{} of the fits are 1.8 for C/O and 2.3 for \omcfe{}. Our quadratic fit for C/O as a function of \feh{} is similar to the quadratic fit derived by \citet{Brewer2016b}. As described in \citet{Brewer2016b}, the metal-poor end of our relation has a similar slope to the linear trends of \citet{Nissen2014} and \citet{Teske2014}, however, the linear trends do not reproduce the leveling off of C/O at higher metallicities.

We can determine \cfe{} and \ofe{} as a function of C/O and \omcfe{}.
\begin{equation}
\begin{split}
    \mathrm{\cfe{}} = \; &\mathomcfe{} - \log_{10}((\mathrm{C/O})^{-1} - 1) \\
    &+ \log_{10}((\mathrm{C/O_\sun})^{-1} - 1) \label{eq:cfe}
\end{split}
\end{equation}
\begin{equation}
\begin{split}
    \mathrm{\ofe{}} = \; &\mathomcfe{} - \log_{10}(1-\mathrm{C/O}) \\
    &+ \log_{10}(1-\mathrm{C/O_\sun}) \label{eq:ofe}
\end{split}
\end{equation}
We do this rather than fit for \cfe{} and \ofe{} directly to better preserve the relation between \feh{} and the relative abundance of C and O, which has a larger effect on the pseudo-continuum than C and O abundances individually. We use these relations to set \cm{} and \om{} of our models based on their \mh{} (using \cm{}, \om{}, and \mh{} as proxies for \cfe{}, \ofe{}, and \feh{}, respectively, in Equations~\ref{eq:ctoo}--\ref{eq:ofe}).

\subsubsection{Treatment of \logg{}}\label{seclogg}

\begin{figure}
\centering
\includegraphics[width=\linewidth]{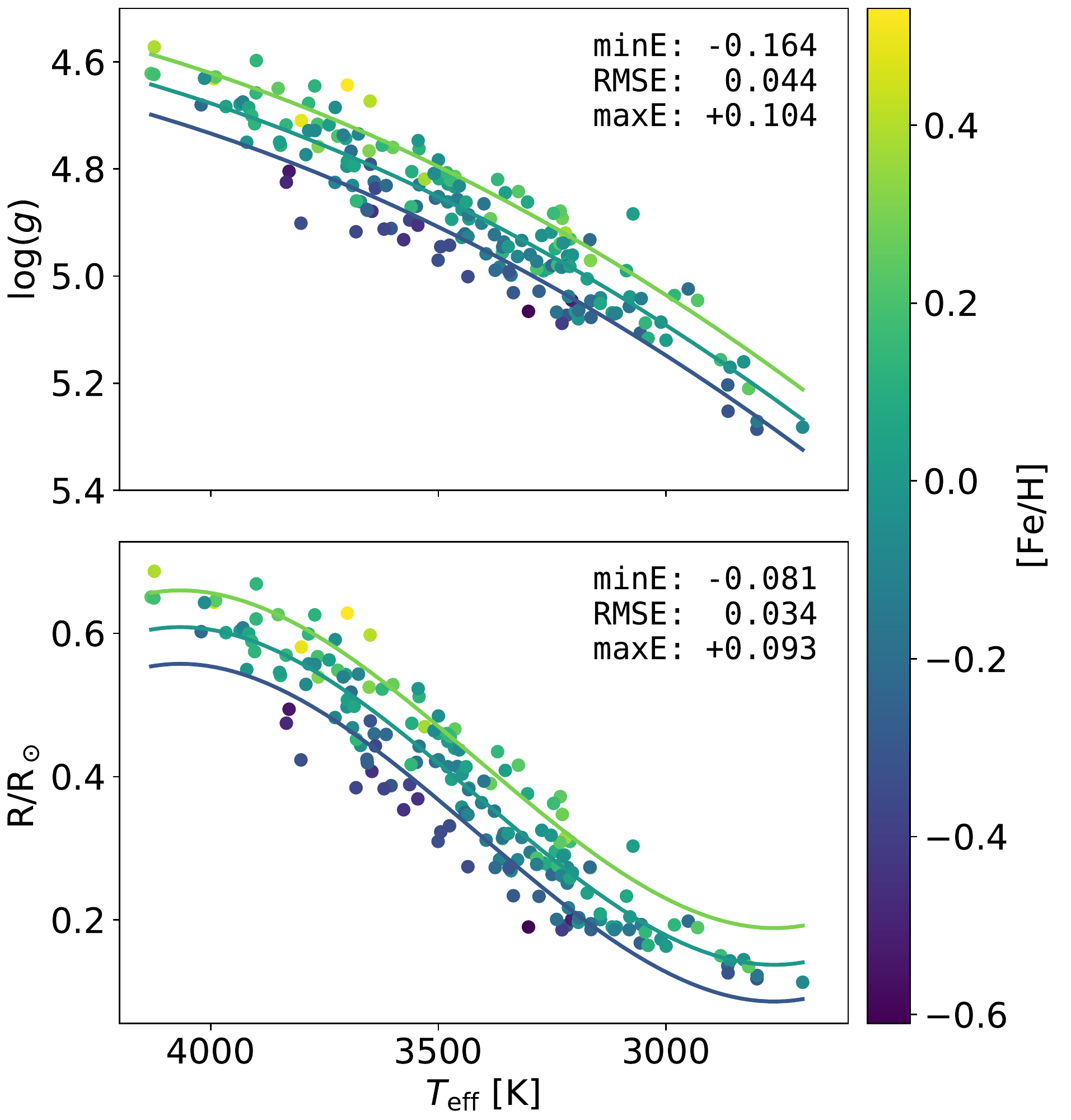}
\caption{\logg{} and radius as a function of \teff{}, colored by \feh{}. Based on data from \citet{Mann2015}. Iso-metallicity fits are shown at \feh{} = $-$0.3, +0.0, and +0.3 based on Equations~\ref{eqlogg}~\&~\ref{eqrad}. \label{teff-logg-rad}}
\end{figure}

All M dwarfs which have reached the main sequence are still on the main sequence, evolving imperceptibly from their ZAMS radius and luminosity \citep{Laughlin1997}. Therefore, we make the assumption that an M dwarf's \logg{} and radius can be determined solely from its temperature and composition. We used the \teff{}, \feh{}, \logg{}, and radius estimates of 183 M dwarfs from \citet{Mann2015} to derive relations for \logg{} and radius as functions of \teff{} and \feh{}. \citet{Mann2015} determined \teff{} by comparing optical spectra to a grid of BT-Settl synthetic spectra, using only spectral regions which resulted in good agreement with effective temperatures derived through LBOI. They calculated radii for their stars from their temperatures and integrated bolometric fluxes via the Stefan-Boltzmann law. They used the calibrations of \citet{Mann2013a} and \citet{Mann2014} to measure metallicities. Originally, \citet{Mann2015} calculated masses from the \citet{Delfosse2000} relation between mass and absolute K-band magnitude. Here, we used the more recent relation of \citet{Benedict2016} to determine masses for use in calculating \logg{}. Figure~\ref{teff-logg-rad} shows \logg{} and radius as a function of \teff{} and \feh{} and our fits to the data. The resulting fits are
\begin{align}
    \begin{split}\label{eqlogg}
        \mlogg{} = & 7.912 - 0.1880 \times \mfeh{} \\
        &- 1.334\mathrm{e}{-3} \times \mteff{} + 1.313\mathrm{e}{-7} \times \mteff{}^2 
    \end{split}\\
    \begin{split}\label{eqrad}
        R_\star/R_\sun = & 15.43 + 0.1708 \times \mfeh{} \\
        & - 1.431\mathrm{e}{-2} \times \mteff{} + 4.350\mathrm{e}{-6} \times \mteff{}^2 \\
        & - 4.246\mathrm{e}{-10} \times \mteff{}^3
    \end{split}
\end{align}
The RMSE of the fits are 0.044 dex in \logg{} and 0.034 $R_\sun$ in radius. Our relation turns over slightly at $\sim$4050 K. While a decrease in radius with increasing temperature is not physical, the \teff{}-radius relation does become more shallow around 4000 K \citep{Boyajian2012}. The absolute difference between the maximum radius and the radius at 4200 K in our relation is less than twice the RMSE in the fit. While the slight turnover is acceptable for our purposes here, we caution against applying these relations beyond the range they are calibrated for. We used these relations to set the \logg{} and radius of our models, removing \logg{} as a major free parameter.

\subsubsection{Model grid sampling}

Following the above simplifications, we are left with three free parameters: \teff{}, \mh{}, and \am{}. Table~\ref{modparams} shows the range and sampling of our grid model parameters.

\begin{deluxetable}{ccc}
\tablecaption{Parameters of the model grid \label{modparams}}
\tablehead{\colhead{Parameter} & \colhead{Range} & \colhead{Step Size}}
\startdata
\teff{} & 3000 -- 4200 K & 100 K \\
\mh{} & $-$1 -- $+$0.5 & 0.25 \\
\am{} & $-$0.1 -- $+$0.4 & 0.1 \\
\enddata
\end{deluxetable}

\begin{deluxetable*}{llcCC}
\tablecaption{Calibration Sample \label{sample}}
\tablehead{
\colhead{M dwarf Name} & \colhead{FGK Name} & \colhead{M dwarf \teff{} [K]} & \colhead{FGK \feh{}} & \colhead{FGK \tife{}}}
\startdata
PM I02441+4913W & HR 799 & 3572 & +0.090 & -0.020 \\
PM I02555+2652 & HD 18143 & 3228 & +0.275 & -0.032 \\
PM I03332+4615S & HIP 16563 & 4075 & +0.079 & -0.017 \\
Gl 166C & HD 26965 & 3167 & -0.290 & +0.220 \\
PM I04559+0440W & HD 31412 & 3570 & +0.110 & -0.010 \\
PM I05415+5329 & HR 1925 & 3765 & +0.150 & -0.050 \\
PM I05463+0112 & HD 38529 & 3642 & +0.350 & -0.030 \\
PM I06461+3233 & HIP 32423 & 3656 & -0.210 & +0.050 \\
PM I07191+6644N & HD 55745 & 4069 & +0.240 & -0.053 \\
PM I08143+6304 & HD 67850 & 3602 & -0.094 & +0.007 \\
PM I08526+2818 & HD 75732 & 3280 & +0.360 & -0.040 \\
PM I09151+2321S & HIP 45406 & 3881 & +0.180 & -0.010 \\
PM I09573+5018 & LSPM J0957+5018E & 3829 & -0.153 & +0.013 \\
PM I11046-0413 & HIP 54155 & 3919 & +0.080 & -0.055 \\
PM I11218+1811 & HIP 55486 & 3993 & +0.358 & -0.066 \\
LSPM J1140+0930E & LSPM J1140+0930W & 3591 & -0.123 & -0.024 \\
PM I13113+0936 & HD 114606 & 4022 & -0.499 & +0.313 \\
PM I13168+1700 & HIP 64797 & 3709 & -0.088 & -0.019 \\
PM I13314-0759W & NLTT 34353 & 3845 & -0.185 & +0.073 \\
LSPM J1404+0157 & LSPM J1404+0156 & 3664 & -0.028 & -0.009 \\
PM I14182+1244W & BD +132777 & 3697 & -0.738 & +0.252 \\
PM I14206-2323N & HIP 70100 & 3950 & +0.178 & +0.007 \\
PM I15118+3933 & HD 135144 & 3435 & -0.076 & +0.006 \\
PM I15164+1647W & HD 135792 & 4106 & -0.284 & +0.107 \\
PM I15204+0011 & HIP 75069 & 3966 & -0.362 & +0.028 \\
PM I16072-1422 & HIP 78969 & 4032 & +0.227 & -0.010 \\
PM I16139+3346 & HD 146362 & 3454 & -0.010 & -0.010 \\
PM I16148+6038 & HD 146868 & 3314 & -0.268 & +0.048 \\
PM I17176+5224 & HIP 84616 & 3231 & -0.071 & +0.017 \\
LSPM J1742+1643 & LSPM J1742+1645 & 3565 & -0.190 & +0.073 \\
PM I17464+2743W & HD 161797 & 3386 & +0.270 & -0.040 \\
PM I18006+6832 & HIP 88188 & 4060 & +0.043 & -0.049 \\
PM I18007+2933 & HD 164595 & 3510 & -0.080 & +0.050 \\
\enddata
\end{deluxetable*}

\subsection{Calibration sample}

We drew our calibration sample from the catalog of common-proper motion FGK+M systems described in \citet{Mann2013a}. In order to empirically calibrate our method, we required accurate measurements of \teff{}, \feh{}, and \tife{} for the M dwarfs in our sample. 

We measured effective temperatures for our M dwarfs using the method described in \citet{Mann2013c} which is also described briefly in Section~\ref{seclogg}. The effective temperatures of these stars were originally calculated along with the sample published in \citet{Mann2015}, although not all stars in this paper were also published there.

We measured the Fe and Ti abundances of our sample from high-resolution optical spectra of the FGK primaries. \citet{Mann2013a} originally obtained and analyzed spectra of the FGK primaries taken with ESPaDOnS on CFHT. Here, we reanalyzed these spectra for 21 stars in our sample using the newest version of SME and following the procedure outlined in \citet{Brewer2016a}. For the other eight stars in our sample, we adopted the abundances derived by \citet{Brewer2016a}. The reported statistical uncertainties from \citet{Brewer2016a} are 0.01 dex in \feh{} and 0.012 dex in [Ti/H]. To ensure consistency between the \citet{Brewer2016a} catalog and abundances measured from ESPaDOnS spectra, we compared abundances for eight stars common to both samples and found they are consistent within measurement errors. We also analyzed three solar spectra reflected from asteroids (two of Ceres, one of Vesta) obtained from the CFHT archive. The derived abundances were consistent with solar abundances to well within measurement uncertainties. For a detailed comparison between abundances derived in this manner and other results from the literature, see \citet{Brewer2016a}.

Table~\ref{sample} lists our calibration sample and their measured properties.

\subsection{Spectral features}

We measure three types of features in our Y-band spectra for use in inferring the \teff{}, [Fe/H], and [Ti/Fe] of M dwarfs: a temperature-sensitive index based on the Wing-Ford FeH band head, the EWs of seven \fei{} lines, and the EWs of ten \tii{} lines. We used the line-identification feature of the \texttt{PHOENIX} models to identify the \ion{Fe}{1} and \ion{Ti}{1} lines and chose wavelength ranges over which to measure EWs based on by-eye inspection of the observed and synthetic spectra. Table~\ref{features} lists the wavelength ranges used when calculating the EWs of these lines and Figure~\ref{spectra} highlights them in a sample of spectra. We define the FeH index as
\begin{equation}
    \mathrm{FeH\ index} = \left\langle F_{\lambda=0.984\textup{--}0.989}\right\rangle / \left\langle F_{\lambda=0.990\textup{--}0.995}\right\rangle,
\end{equation}
where $\left\langle F_{\lambda=a\textup{--}b}\right\rangle$ is the mean flux in the interval $\lambda=a\textup{--}b$. As shown in Figure~\ref{spectra} the strength of the FeH band head has a strong spectral-type dependence and is deeper in later M dwarfs. Being a Fe-bearing molecule, it is also sensitive to [Fe/H], but to a lesser extent. The temperature-sensitivity of FeH lines has been noted in previous works \citep[e.g.,][]{Onehag2012}.

\begin{deluxetable*}{ccDDDD}
\tablecaption{Y-band features \label{features}}
\tablewidth{0pt}
\tablehead{
\colhead{Feature} & \colhead{Wavelength Range [$\micron$]} & \multicolumn2c{$a_1$\tablenotemark{a}} & \multicolumn2c{$a_2$\tablenotemark{a}} & \multicolumn2c{$a_3$\tablenotemark{a}} & \multicolumn2c{RMSE [$\AA$]}}
\decimals
\startdata
FeH index & 0.984--0.989, 0.990--0.995 & -0.0574 & 1.07 & .  & 0.00355 \\
\fei{} line & 1.01475--1.01506 & -1.48 & 0.686 & 1.49 & 0.0143 \\
\fei{} line & 1.02183--1.02200 & -1.24 & 0.582 & 1.22 & 0.00779 \\
\fei{} line & 1.03980--1.03990 & -0.698 & 0.729 & 0.695 & 0.00498 \\
\fei{} line & 1.04253--1.04273 & -1.57 & 0.67 & 1.56 & 0.0128 \\
\fei{} line & 1.04719--1.04733 & -0.167 & 0.823 & 0.164 & 0.00702 \\
\fei{} line & 1.05343--1.05360 & 0.08 & 0.89 & -0.0925 & 0.00647 \\
\fei{} line & 1.07854--1.07867 & -0.172 & 1.05 & 0.156 & 0.00776 \\
\tii{} line & 1.00001--1.00013 & -0.773 & 0.575 & 0.777 & 0.00777 \\
\tii{} line & 1.00367--1.00378 & -0.751 & 0.633 & 0.755 & 0.00556 \\
\tii{} line & 1.00597--1.00609 & -0.287 & 0.75 & 0.294 & 0.00532 \\
\tii{} line & 1.03990--1.04009 & -1.56 & 0.755 & 1.56 & 0.0143 \\
\tii{} line & 1.04979--1.05000 & -0.34 & 1.04 & 0.278 & 0.0129 \\
\tii{} line & 1.05866--1.05886 & -1.48 & 0.662 & 1.47 & 0.0154 \\
\tii{} line & 1.06100--1.06111 & -0.46 & 0.568 & 0.462 & 0.00471 \\
\tii{} line & 1.06793--1.06806 & -0.799 & 0.604 & 0.806 & 0.0063 \\
\tii{} line & 1.07285--1.07300 & -0.955 & 0.817 & 0.954 & 0.00909 \\
\tii{} line & 1.07768--1.07787 & -0.878 & 0.353 & 0.903 & 0.00784 \\
\enddata
\tablenotetext{a}{$a_\text{1--2}$ for FeH index, $b_\text{1--3}$ for \fei{} lines, $c_\text{1--3}$ for \tii{} lines}
\end{deluxetable*}

\subsubsection{Determining the pseudo-continuum level}

Defining the continuum level in an M dwarf's spectrum is a long-standing problem for M dwarf abundance analysis. One commonly implemented solution is to choose two ``continuum regions'' on either side of the feature that are relatively free of absorption features and linearly interpolate between the mean flux in the two regions. However, these regions can be small and significantly effected by statistical (photon noise) or systematic (e.g., variations in molecular opacity or poor telluric correction) variations. To mitigate these issues, we developed a new method of assigning the pseudo-continuum level that is less sensitive to anomalous data points and can consistently assess the pseudo-continuum across different targets and spectral regions. Since we correct for the instrument throughput and place all Echelle orders on the same relative scale, we can assign the pseudo-continuum over the entire Y-band at once (orders 71--77, 0.98--1.08 $\micron$). The process has three steps. First, we use a 2nd order Savitzky–Golay filter \citep{Savitzky1964} with a window length of five pixels to reduce high frequency variations in the spectrum. Then, we apply a running maximum filter with a width of seven resolution elements. Finally, we fit a 6th order Chebyshev polynomial to the filtered spectrum to use as the pseudo-continuum. Figure~\ref{spectra} shows examples of the continuum fits.

We list all indices and EWs measured from the NIRSPEC spectra of our calibration sample in Table~\ref{sample_ews}. We list all indices and EWs measured from our model grid in Table~\ref{grid_ews}.

\begin{figure*}
\centering
\includegraphics[width=\linewidth]{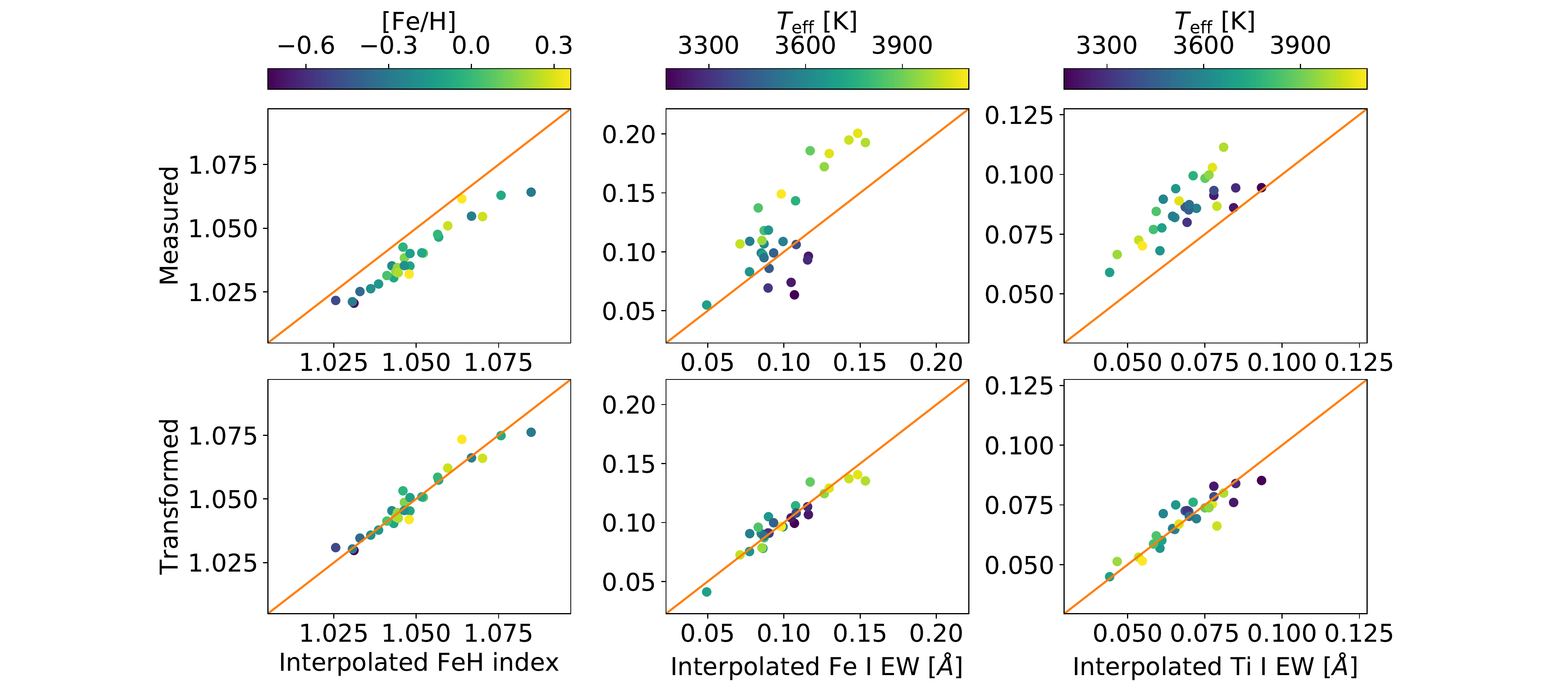}
\caption{The FeH index, a \fei{} EW, and a \tii{} EW as interpolated from our model grid compared with those measured directly from our NIRSPEC spectra (top) and the measured values transformed as described in the text (bottom). An orange line denotes the 1:1 relation. The models do not correctly predict the observed line strengths, but simple transformations of the observed indices and EWs are enough to bring the observations and models into agreement. \label{ews_comp}}
\end{figure*}

\subsection{Calibrating the models}

If we could fully trust the synthetic M dwarf spectra, we could generate spectra for the known parameters of each star and use curve of growth analysis to determine abundances. To test the agreement between our Y-band models and our NIRSPEC observations, we show in the top row of Figure~\ref{ews_comp} an example comparison of line EWs measured from our NIRSPEC spectra and EWs interpolated from our model grid based on the known parameters of our calibration sample. The EWs and indices predicted by the models are close to, but not a perfect match to what we measure from our NIRSPEC spectra.

The fact that we know the temperatures and compositions of these systems beforehand allows use to empirically derive simple transformations of the observed indices and EWs in order to force agreement between the models and observations. After analyzing the relations between line EWs measured from our NIRSPEC spectra and EWs interpolated from our model grid based on known parameters (examples shown in the top row of Figure~\ref{ews_comp}), we formulated the following transformations to the observed features.
\begin{alignat}{3}
    I_\mathrm{FeH}' & = a_1 &&+ a_2 I_\mathrm{FeH} \label{eq:trans_feh}\\
    \mathrm{EW}'_\mathrm{Fe} & = b_1 &&+ b_2 \mathrm{EW}_\mathrm{Fe} &&+ b_3 I_\mathrm{FeH} \label{eq:trans_feew}\\
    \mathrm{EW}'_\mathrm{Ti} & = c_1 &&+ c_2 \mathrm{EW}_\mathrm{Ti} &&+ c_3 I_\mathrm{FeH} \label{eq:trans_tiew}
\end{alignat}
Here, $I_\mathrm{FeH}$, $\mathrm{EW}_\mathrm{Fe}$, and $\mathrm{EW}_\mathrm{Ti}$ denote the FeH index, \fei{} line EW, and \tii{} line EW, respectively, measured from a NIRSPEC spectrum. Primed values are the transformed index and EWs. By relying only on values measurable directly from the NIRSPEC spectra, these transformations do not require any prior knowledge of stellar parameters.

\begin{figure*}
\centering
\includegraphics[width=\linewidth]{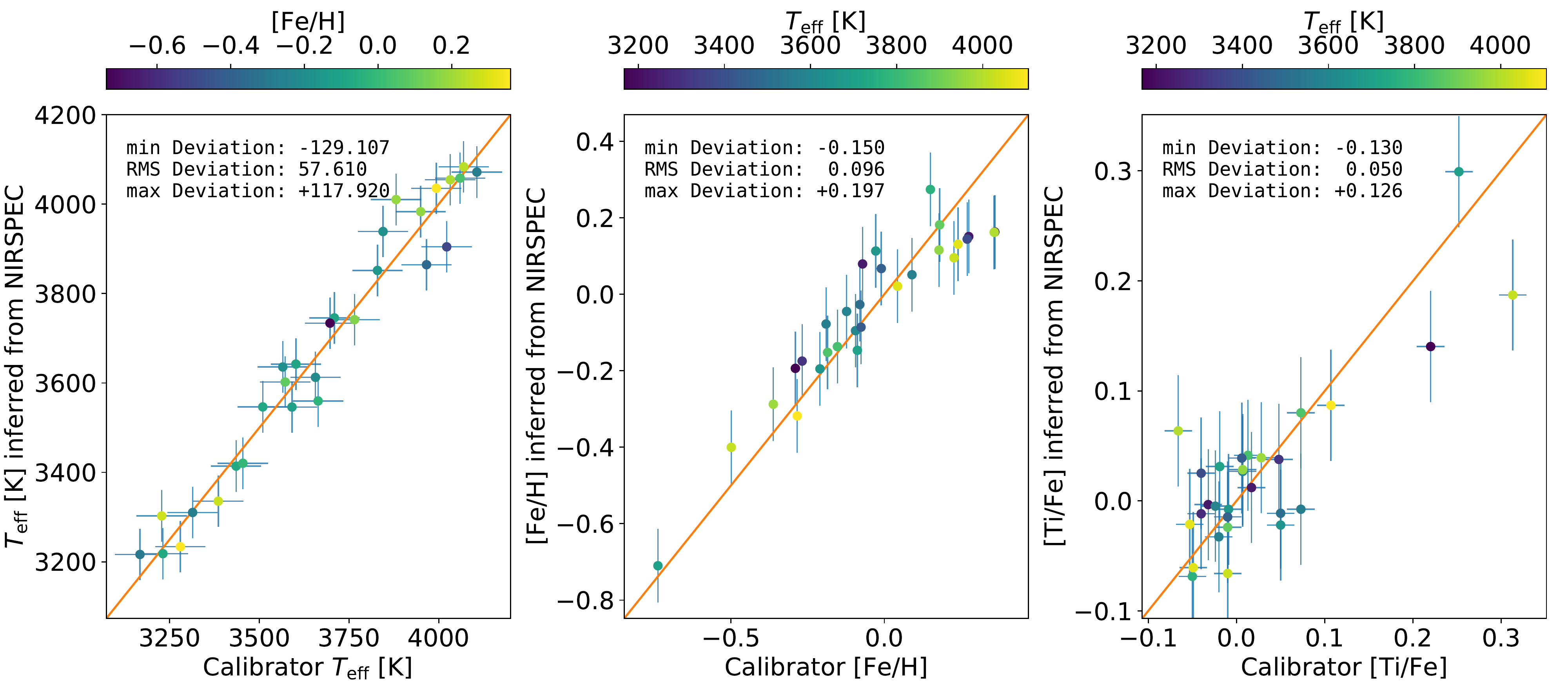}
\caption{Comparison between known \teff{}, \feh{}, and \tife{} of the calibration sample and those parameters inferred from our NIRSPEC spectra as described in Section~\ref{method}. An orange line denotes the 1:1 relation. X error bars show measurement error in the known parameters. Y error bars indicate the RMSE between known and inferred parameters. Minimum, RMS, and maximum deviations from known parameter values are shown for each parameter. \label{fit_vs_real_params}}
\end{figure*}

For each spectral feature (Table~\ref{features}), we fit for the $a_\text{1--2}$, $b_\text{1--3}$, or $c_\text{1--3}$ constants via least squares between the indices or EWs interpolated from our model grid and the indices or EWs measured from our NIRSPEC spectra and transformed following Equations~\ref{eq:trans_feh}--\ref{eq:trans_tiew}. Table~\ref{features} lists the best fit constants and the RMSE in the residuals. The bottom row of Figure~\ref{ews_comp} shows examples of the transformed measurements.

\subsection{A method to measure \teff{}, \feh{}, and \tife{}}\label{method}

The calibration described above can be inverted to determine \teff{}, \feh{}, and \tife{} of a star based on features measured from its Y-band spectrum. First, the measured FeH index, \fei{} EWs, and \tii{} EWs are transformed via Equations~\ref{eq:trans_feh}--\ref{eq:trans_feew} with constants from Table~\ref{features}. Then, best fit \teff{}, \feh{}, and \tife{} are determined via \chisq{} minimization. \chisq{} is calculated as
\begin{equation}
    \chi^2 = \sum_i \left(\frac{f'_i - \hat{f}_i\left(\mteff{},\mfeh{},\mtife{}\right)}{\sigma_{f_i}}\right)^2,
\end{equation}
where $i$ indicates the $i$th feature in Table~\ref{features}, $f'_i$ is the transformed index or EW, $\hat{f}_i\left(\mteff{},\mfeh{},\mtife{}\right)$ is the index or EW interpolated from the model grid based on the fitted parameters, and $\sigma_i$ is the RMSE of the residuals from the transformation calibration (last column of Table~\ref{features}). This assumes that residuals in the transformation dominate over EW uncertainty due to photon noise.

To estimate the uncertainty in the inferred parameters due to the inherent uncertainty of the above procedure, we applied it to our calibration sample and compared with the known parameters. The results are shown in Figure~\ref{fit_vs_real_params}. We achieve internal precisions of 60 K, 0.1 dex, and 0.05 dex in \teff{}, \feh{}, and \tife{}, respectively.

We have made code to estimate \teff{}, \feh{}, and \tife{} of an M dwarf from its NIRSPEC Y-band spectrum publicly available at \url{https://github.com/mveyette/analyze_NIRSPEC1}. The code is written in Python 3 and performs throughput correction, assigns the pseudo-continuum, measures EWs and indices, applies our empirical corrections, and matches to our model grid to return the best-fit \teff{}, \feh{}, and \tife{}.

\section{Discussion}\label{disc}
By measuring Fe and Ti abundances directly from \ion{Fe}{1} and \ion{Ti}{1} lines, we improve upon previous empirical metallicity calibrations that rely on correlated abundance trends. Furthermore, we do not have to assume the nonlinear functional form for how line EWs change as a function of stellar parameters. Instead, we leverage the complex physical prescription of the \texttt{PHOENIX} models to account for most of the change and apply simple, easily determined corrections to force agreement with our calibration sample. 

Our method does, however, suffer some drawbacks. In order to create a generalized grid of synthetic spectra for inferring \teff{}, \feh{}, and \tife{}, we had to make some assumptions regarding other physical properties of M dwarfs. The underlying assumption is that all main-sequence M dwarfs can be uniquely characterized by their \teff{}, \feh{}, and \afe{} alone. In reality, this is not the case.

In terms of other abundances, C and O have the largest effect on an M dwarf's spectrum \citep{Veyette2016b}. While, most stars around solar metallicity fall within a narrow range of C/O and \omcfe{}, there is likely still some inherent scatter in the \feh{}-C/O and \feh{}-\omcfe{} relations that is not captured in our analysis. Furthermore, very metal-poor stars (\feh{} $<$ $-$0.75) show a larger spread in C/O and \omcfe{}, separable as a low-alpha halo group (O-poor) and a high-alpha halo/thick disk group (O-rich). This is more evident in the results of \citet{Nissen2014} as they analyzed more metal-poor stars than \citet{Brewer2016a}. This intrinsic scatter in C and O introduces additional uncertainty in our method. We note that there are only a few stars with \feh{}~$<$~$-$0.5 in the \citet{Brewer2016a} sample, which, combined with the fact that our FGK+M calibration sample only contains two stars with \feh{}~$\le$~$-$0.5, means our calibration is poorly constrained in the low-metallicity regime.

By assuming the entire composition of a star can be parameterized solely by \feh{} and \afe{}, our method will likely fail for stars with non-standard abundance ratios such as stars that have accreted processed material from an evolved companion \citep[e.g., dwarf carbon stars,][]{Green2013}.

There exists a slight systematic trend in the residuals between the calibration sample \feh{} and the \feh{} we estimate from the NIRSPEC spectra (Figure~\ref{fit_vs_real_params}). It is unclear what the exact origin of this trend is, though it is likely that it is residual systematic differences between the models and our observations that are not accounted for by our simple corrections to the EWs. The mean of the residuals is $<$0.003 dex, however, sub-solar metallicity stars tend to be overestimated in \feh{} while super-solar metallicity stars tend to be underestimated. This may suggest a tendency to favor solar metallicity models. If we fit for and remove this residual trend, the RMSE is reduced to 0.06 dex.

This work highlights an important limitation of current low-mass star synthetic spectra. Differences between observed and model line strengths can be due to many different issues. Some of these issues are more significant for M dwarfs than for FGK stars, others are unique to M dwarfs. The corrective transformation we apply to our measured EWs and indices is meant to account for the combined effect of these issues. Here we list some potential issues.

Inaccurate oscillator strengths of the lines we used in this analysis could be a major reason why observed line strengths do not match modeled line strengths. We did not attempt to adjust the oscillator strengths of any lines used in this analysis.

The thermal profile of the stellar atmosphere model used plays a large role in determining the flux inside individual absorption lines as well as of the pseudo-continuum level from which line strengths are measured. Incomplete or inaccurate line lists for major opacity sources can result in an inaccurate equilibrium thermal profile. This is particularly an issue for M dwarfs as nearly all the flux emitted by M dwarf is emitted at wavelengths where there is at least some molecular opacity. Current line lists for major opacity sources in M dwarf atmospheres such as TiO are known to be inaccurate and incomplete \citep[e.g.,]{Rajpurohit2013,Mann2013c,Hoeijmakers2015}. However, new advances in experimental and theoretical studies of rotational-vibrational energy levels for important molecules, largely motivated by their application to observations of exoplanet atmospheres, may improve future cool dwarf models \citep[e.g.,][]{Tennyson2016,McKemmish2017}.

Other inaccuracies in model parameters may play small roles in the overall mismatch between synthetic spectra and observations, such as: mixing length, determined from the \teff{} and \logg{} of the model according to the calibration of \citet{Ludwig1999}; microturbulent velocity, determined from \teff{} according to the radiation hydrodynamic simulations of \citet{Freytag2010}; and \logg{}, determined from \teff{} and \mh{} as described in Section~\ref{seclogg}. 

One effect not accounted for in the \texttt{PHOENIX} models is line splitting in the presence of magnetic fields. Some FeH lines in the Wing-Ford band are magnetically sensitive \citep{Reiners2007}. Varying magnetic field strengths may introduce additional uncertainty to our method. However, strong magnetic fields are associated with rapid rotation \citep{Noyes1984}. Stars with strong enough magnetic fields to significantly effect our FeH index will already be excluded from our analysis due to significant rotational broadening. We note that \citet{Shulyak2014} found the \ion{Ti}{1} lines in Y band are not very magnetically sensitive.

\section{Concluding Remarks}\label{summary}
We developed a method to measure \teff{}, \feh{}, and \tife{} from high-resolution Y-band spectra of M dwarfs. Our method is physically motivated in that it relies on state-of-the-art stellar atmosphere models to provide the nonlinear relations for how M dwarf spectra change as a function of temperature and composition. Our method is also empirically calibrated, using observations of M dwarfs with wide FGK companions to force agreement between known parameters and those inferred from our NIR spectra. Unlike other empirical metallicity calibrations, our method measures Fe and Ti abundances directly from atomic \fei{} and \tii{} lines. Our calibration achieves precisions in \teff{}, \feh{}, and \tife{} of 60 K, 0.1 dex, and 0.05 dex, respectively. Improvements to cool dwarf atmosphere models and larger calibration samples with wider wavelength coverage could allow for detailed chemical analysis of M dwarfs at a similar precision achieved for FGK stars.

Few high-resolution, Y-band spectrometers are currently available, which limits the application of the method presented here. However, at least three new exoplanet RV surveys specifically targeting M dwarfs include coverage of Y band: CARMENES \citep{Quirrenbach2010}, the Habitable Zone Planet Finder \citep[HPF,][]{Mahadevan2010}, and SPIRou \citep{Artigau2011}. These surveys will provide high-S/N, high-resolution, Y-band spectra for hundreds of M dwarfs, many of which host planets that will be detected during the surveys. The ability to detect planets around and measure \feh{} and \tife{} for hundreds of M dwarfs using the same dataset will be a powerful asset. These surveys will allow us to test whether observed trends in the composition of planet-hosting FGK stars, like enhanced \al{}-element abundance, also hold for lower mass stars and smaller planets.

Another exciting application of this work is the potential to use alpha-enhancement to constrain ages of field M dwarfs. The age of an individual field M dwarf is difficult to measure reliably because its radius and effective temperature change imperceptibly once on the main sequence \citep{Laughlin1997}. However,  surveys of nearby stars find an empirical relation between \afe{} and age \citep{Haywood2013, Bensby2014, Feuillet2016} due to delayed Fe enrichment of the ISM by type Ia supernovae. Measuring \afe{} of an M dwarf can be combined with priors based on kinematics \citep[e.g.,][]{Burgasser2017} to provide a powerful age diagnostic. The ability to measure ages of field M dwarfs has many applications including constraining the age-rotation-activity relation of M dwarfs \citep[e.g.,][]{Newton2016,Newton2017}.

\acknowledgments

We thank the anonymous referee for their thoughtful comments and useful suggestions.

This material is based upon work supported by the National Science Foundation under Grant No. 1716260.  Additional support for this work was provided by a NASA Keck PI Data Award, administered by the NASA Exoplanet Science Institutem and by the Department of Astronomy and the Institute for Astrophysical Research at Boston University. A.W.M. is supported through Hubble Fellowship grant 51364 awarded by the Space Telescope Science Institute, which is operated by the Association of Universities for Research in Astronomy, Inc., for NASA, under contract NAS 5-26555. F.A. received funding from the French ``Programme National de Physique Stellaire'' (PNPS) and the ``Programme National de Planetologie'' of CNRS (INSU). D.H. is supported by the Collaborative Research Centre SFB 881 ``The Milky Way System'' (subproject A4) of the German Research Foundation (DFG).

This research made use of the Massachusetts Green High Performance Computing Center in Holyoke, MA.

The data presented herein were obtained at the W.M. Keck Observatory, which is operated as a scientific partnership among the California Institute of Technology, the University of California and the National Aeronautics and Space Administration. The Observatory was made possible by the generous financial support of the W.M. Keck Foundation.

The authors wish to recognize and acknowledge the very significant cultural role and reverence that the summit of Mauna Kea has always had within the indigenous Hawaiian community.  We are most fortunate to have the opportunity to conduct observations from this mountain.

\bibliographystyle{apj}
\bibliography{bib}

\begin{thebibliography}{}
\expandafter\ifx\csname natexlab\endcsname\relax\def\natexlab#1{#1}\fi

\bibitem[{{Adibekyan} {et~al.}(2012{\natexlab{a}}){Adibekyan}, {Sousa},
  {Santos}, {Delgado Mena}, {Gonz{\'a}lez Hern{\'a}ndez}, {Israelian}, {Mayor},
  \& {Khachatryan}}]{Adibekyan2012b}
{Adibekyan}, V.~Z., {Sousa}, S.~G., {Santos}, N.~C., {et~al.}
  2012{\natexlab{a}}, \aap, 545, A32

\bibitem[{{Adibekyan} {et~al.}(2012{\natexlab{b}}){Adibekyan}, {Santos},
  {Sousa}, {Israelian}, {Delgado Mena}, {Gonz{\'a}lez Hern{\'a}ndez}, {Mayor},
  {Lovis}, \& {Udry}}]{Adibekyan2012a}
{Adibekyan}, V.~Z., {Santos}, N.~C., {Sousa}, S.~G., {et~al.}
  2012{\natexlab{b}}, \aap, 543, A89

\bibitem[{{Allard}(2016)}]{Allard2016}
{Allard}, F. 2016, in SF2A-2016: Proceedings of the Annual meeting of the
  French Society of Astronomy and Astrophysics, ed. C.~{Reyl{\'e}},
  J.~{Richard}, L.~{Cambr{\'e}sy}, M.~{Deleuil}, E.~{P{\'e}contal},
  L.~{Tresse}, \& I.~{Vauglin}, 223--227

\bibitem[{{Allard} {et~al.}(2012){Allard}, {Homeier}, \&
  {Freytag}}]{Allard2012a}
{Allard}, F., {Homeier}, D., \& {Freytag}, B. 2012, Royal Society of London
  Philosophical Transactions Series A, 370, 2765

\bibitem[{{Allard} {et~al.}(2013){Allard}, {Homeier}, {Freytag},
  {Schaffenberger}, {}, \& {Rajpurohit}}]{Allard2013}
{Allard}, F., {Homeier}, D., {Freytag}, B., {et~al.} 2013, Memorie della
  Societa Astronomica Italiana Supplementi, 24, 128

\bibitem[{{Alvarez} \& {Plez}(1998)}]{Alvarez1998}
{Alvarez}, R., \& {Plez}, B. 1998, \aap, 330, 1109

\bibitem[{{Artigau} {et~al.}(2011){Artigau}, {Donati}, \&
  {Delfosse}}]{Artigau2011}
{Artigau}, {\'E}., {Donati}, J.-F., \& {Delfosse}, X. 2011, in Astronomical
  Society of the Pacific Conference Series, Vol. 448, 16th Cambridge Workshop
  on Cool Stars, Stellar Systems, and the Sun, ed. C.~{Johns-Krull}, M.~K.
  {Browning}, \& A.~A. {West}, 771

\bibitem[{{Asplund} {et~al.}(2009){Asplund}, {Grevesse}, {Sauval}, \&
  {Scott}}]{Asplund2009}
{Asplund}, M., {Grevesse}, N., {Sauval}, A.~J., \& {Scott}, P. 2009, \araa, 47,
  481

\bibitem[{{Astudillo-Defru} {et~al.}(2017){Astudillo-Defru}, {Forveille},
  {Bonfils}, {S{\'e}gransan}, {Bouchy}, {Delfosse}, {Lovis}, {Mayor}, {Murgas},
  {Pepe}, {Santos}, {Udry}, \& {W{\"u}nsche}}]{Astudillo-Defru2017}
{Astudillo-Defru}, N., {Forveille}, T., {Bonfils}, X., {et~al.} 2017, \aap,
  602, A88

\bibitem[{{Baraffe} {et~al.}(2015){Baraffe}, {Homeier}, {Allard}, \&
  {Chabrier}}]{Baraffe2015}
{Baraffe}, I., {Homeier}, D., {Allard}, F., \& {Chabrier}, G. 2015, \aap, 577,
  A42

\bibitem[{{Bean} {et~al.}(2006){Bean}, {Sneden}, {Hauschildt}, {Johns-Krull},
  \& {Benedict}}]{Bean2006}
{Bean}, J.~L., {Sneden}, C., {Hauschildt}, P.~H., {Johns-Krull}, C.~M., \&
  {Benedict}, G.~F. 2006, \apj, 652, 1604

\bibitem[{{Benedict} {et~al.}(2016){Benedict}, {Henry}, {Franz}, {McArthur},
  {Wasserman}, {Jao}, {Cargile}, {Dieterich}, {Bradley}, {Nelan}, \&
  {Whipple}}]{Benedict2016}
{Benedict}, G.~F., {Henry}, T.~J., {Franz}, O.~G., {et~al.} 2016, \aj, 152, 141

\bibitem[{{Bensby} {et~al.}(2014){Bensby}, {Feltzing}, \& {Oey}}]{Bensby2014}
{Bensby}, T., {Feltzing}, S., \& {Oey}, M.~S. 2014, \aap, 562, A71

\bibitem[{{Bonfils} {et~al.}(2005){Bonfils}, {Delfosse}, {Udry}, {Santos},
  {Forveille}, \& {S{\'e}gransan}}]{Bonfils2005}
{Bonfils}, X., {Delfosse}, X., {Udry}, S., {et~al.} 2005, \aap, 442, 635

\bibitem[{{Boyajian} {et~al.}(2012){Boyajian}, {von Braun}, {van Belle},
  {McAlister}, {ten Brummelaar}, {Kane}, {Muirhead}, {Jones}, {White},
  {Schaefer}, {Ciardi}, {Henry}, {L{\'o}pez-Morales}, {Ridgway}, {Gies}, {Jao},
  {Rojas-Ayala}, {Parks}, {Sturmann}, {Sturmann}, {Turner}, {Farrington},
  {Goldfinger}, \& {Berger}}]{Boyajian2012}
{Boyajian}, T.~S., {von Braun}, K., {van Belle}, G., {et~al.} 2012, \apj, 757,
  112

\bibitem[{{Brewer} \& {Fischer}(2016)}]{Brewer2016b}
{Brewer}, J.~M., \& {Fischer}, D.~A. 2016, \apj, 831, 20

\bibitem[{{Brewer} {et~al.}(2016){Brewer}, {Fischer}, {Valenti}, \&
  {Piskunov}}]{Brewer2016a}
{Brewer}, J.~M., {Fischer}, D.~A., {Valenti}, J.~A., \& {Piskunov}, N. 2016,
  \apjs, 225, 32

\bibitem[{{Brugamyer} {et~al.}(2011){Brugamyer}, {Dodson-Robinson}, {Cochran},
  \& {Sneden}}]{Brugamyer2011}
{Brugamyer}, E., {Dodson-Robinson}, S.~E., {Cochran}, W.~D., \& {Sneden}, C.
  2011, \apj, 738, 97

\bibitem[{{Buchhave} \& {Latham}(2015)}]{Buchhave2015}
{Buchhave}, L.~A., \& {Latham}, D.~W. 2015, \apj, 808, 187

\bibitem[{{Buchhave} {et~al.}(2012){Buchhave}, {Latham}, {Johansen},
  {Bizzarro}, {Torres}, {Rowe}, {Batalha}, {Borucki}, {Brugamyer}, {Caldwell},
  {Bryson}, {Ciardi}, {Cochran}, {Endl}, {Esquerdo}, {Ford}, {Geary},
  {Gilliland}, {Hansen}, {Isaacson}, {Laird}, {Lucas}, {Marcy}, {Morse},
  {Robertson}, {Shporer}, {Stefanik}, {Still}, \& {Quinn}}]{Buchhave2012}
{Buchhave}, L.~A., {Latham}, D.~W., {Johansen}, A., {et~al.} 2012, \nat, 486,
  375

\bibitem[{{Buchhave} {et~al.}(2014){Buchhave}, {Bizzarro}, {Latham},
  {Sasselov}, {Cochran}, {Endl}, {Isaacson}, {Juncher}, \&
  {Marcy}}]{Buchhave2014}
{Buchhave}, L.~A., {Bizzarro}, M., {Latham}, D.~W., {et~al.} 2014, \nat, 509,
  593

\bibitem[{{Burgasser} \& {Mamajek}(2017)}]{Burgasser2017}
{Burgasser}, A.~J., \& {Mamajek}, E.~E. 2017, \apj, 845, 110

\bibitem[{{Caffau} {et~al.}(2011){Caffau}, {Ludwig}, {Steffen}, {Freytag}, \&
  {Bonifacio}}]{Caffau2011}
{Caffau}, E., {Ludwig}, H.-G., {Steffen}, M., {Freytag}, B., \& {Bonifacio}, P.
  2011, \solphys, 268, 255

\bibitem[{{Casagrande} {et~al.}(2008){Casagrande}, {Flynn}, \&
  {Bessell}}]{Casagrande2008}
{Casagrande}, L., {Flynn}, C., \& {Bessell}, M. 2008, \mnras, 389, 585

\bibitem[{{Cushing} {et~al.}(2004){Cushing}, {Vacca}, \&
  {Rayner}}]{Cushing2004}
{Cushing}, M.~C., {Vacca}, W.~D., \& {Rayner}, J.~T. 2004, \pasp, 116, 362

\bibitem[{{Delfosse} {et~al.}(2000){Delfosse}, {Forveille}, {S{\'e}gransan},
  {Beuzit}, {Udry}, {Perrier}, \& {Mayor}}]{Delfosse2000}
{Delfosse}, X., {Forveille}, T., {S{\'e}gransan}, D., {et~al.} 2000, \aap, 364,
  217

\bibitem[{{Delgado Mena} {et~al.}(2010){Delgado Mena}, {Israelian},
  {Gonz{\'a}lez Hern{\'a}ndez}, {Bond}, {Santos}, {Udry}, \&
  {Mayor}}]{Delgado2010}
{Delgado Mena}, E., {Israelian}, G., {Gonz{\'a}lez Hern{\'a}ndez}, J.~I.,
  {et~al.} 2010, \apj, 725, 2349

\bibitem[{{Dittmann} {et~al.}(2016){Dittmann}, {Irwin}, {Charbonneau}, \&
  {Newton}}]{Dittman2016}
{Dittmann}, J.~A., {Irwin}, J.~M., {Charbonneau}, D., \& {Newton}, E.~R. 2016,
  \apj, 818, 153

\bibitem[{{Everett} {et~al.}(2013){Everett}, {Howell}, {Silva}, \&
  {Szkody}}]{Everett2013}
{Everett}, M.~E., {Howell}, S.~B., {Silva}, D.~R., \& {Szkody}, P. 2013, \apj,
  771, 107

\bibitem[{{Feuillet} {et~al.}(2016){Feuillet}, {Bovy}, {Holtzman}, {Girardi},
  {MacDonald}, {Majewski}, \& {Nidever}}]{Feuillet2016}
{Feuillet}, D.~K., {Bovy}, J., {Holtzman}, J., {et~al.} 2016, \apj, 817, 40

\bibitem[{{Fischer} \& {Valenti}(2005)}]{Fischer2005}
{Fischer}, D.~A., \& {Valenti}, J. 2005, \apj, 622, 1102

\bibitem[{{Freytag} {et~al.}(2010){Freytag}, {Allard}, {Ludwig}, {Homeier}, \&
  {Steffen}}]{Freytag2010}
{Freytag}, B., {Allard}, F., {Ludwig}, H.-G., {Homeier}, D., \& {Steffen}, M.
  2010, \aap, 513, A19

\bibitem[{{Gaidos} {et~al.}(2016){Gaidos}, {Mann}, {Kraus}, \&
  {Ireland}}]{Gaidos2016}
{Gaidos}, E., {Mann}, A.~W., {Kraus}, A.~L., \& {Ireland}, M. 2016, \mnras,
  457, 2877

\bibitem[{{Gonzalez}(1997)}]{Gonzalez1997}
{Gonzalez}, G. 1997, \mnras, 285, 403

\bibitem[{{Green}(2013)}]{Green2013}
{Green}, P. 2013, \apj, 765, 12

\bibitem[{{Gustafsson} {et~al.}(2008){Gustafsson}, {Edvardsson}, {Eriksson},
  {J{\o}rgensen}, {Nordlund}, \& {Plez}}]{Gustafsson2008}
{Gustafsson}, B., {Edvardsson}, B., {Eriksson}, K., {et~al.} 2008, \aap, 486,
  951

\bibitem[{{Haywood} {et~al.}(2013){Haywood}, {Di Matteo}, {Lehnert}, {Katz}, \&
  {G{\'o}mez}}]{Haywood2013}
{Haywood}, M., {Di Matteo}, P., {Lehnert}, M.~D., {Katz}, D., \& {G{\'o}mez},
  A. 2013, \aap, 560, A109

\bibitem[{{Hejazi} {et~al.}(2015){Hejazi}, {De Robertis}, \&
  {Dawson}}]{Hejazi2015}
{Hejazi}, N., {De Robertis}, M.~M., \& {Dawson}, P.~C. 2015, \aj, 149, 140

\bibitem[{{Hoeijmakers} {et~al.}(2015){Hoeijmakers}, {de Kok}, {Snellen},
  {Brogi}, {Birkby}, \& {Schwarz}}]{Hoeijmakers2015}
{Hoeijmakers}, H.~J., {de Kok}, R.~J., {Snellen}, I.~A.~G., {et~al.} 2015,
  \aap, 575, A20

\bibitem[{{Horne}(1986)}]{Horne1986}
{Horne}, K. 1986, \pasp, 98, 609

\bibitem[{{Johnson} {et~al.}(2010){Johnson}, {Aller}, {Howard}, \&
  {Crepp}}]{Johnson2010}
{Johnson}, J.~A., {Aller}, K.~M., {Howard}, A.~W., \& {Crepp}, J.~R. 2010,
  \pasp, 122, 905

\bibitem[{{Johnson} \& {Apps}(2009)}]{Johnson2009}
{Johnson}, J.~A., \& {Apps}, K. 2009, \apj, 699, 933

\bibitem[{{Johnson} {et~al.}(2012){Johnson}, {Gazak}, {Apps}, {Muirhead},
  {Crepp}, {Crossfield}, {Boyajian}, {von Braun}, {Rojas-Ayala}, {Howard},
  {Covey}, {Schlawin}, {Hamren}, {Morton}, {Marcy}, \& {Lloyd}}]{Johnson2012}
{Johnson}, J.~A., {Gazak}, J.~Z., {Apps}, K., {et~al.} 2012, \aj, 143, 111

\bibitem[{{Laughlin} {et~al.}(1997){Laughlin}, {Bodenheimer}, \&
  {Adams}}]{Laughlin1997}
{Laughlin}, G., {Bodenheimer}, P., \& {Adams}, F.~C. 1997, \apj, 482, 420

\bibitem[{{Lindgren} \& {Heiter}(2017)}]{Lindgren2017}
{Lindgren}, S., \& {Heiter}, U. 2017, \aap, 604, A97

\bibitem[{{Lindgren} {et~al.}(2016){Lindgren}, {Heiter}, \&
  {Seifahrt}}]{Lindgren2016}
{Lindgren}, S., {Heiter}, U., \& {Seifahrt}, A. 2016, \aap, 586, A100

\bibitem[{{Ludwig} {et~al.}(1999){Ludwig}, {Freytag}, \&
  {Steffen}}]{Ludwig1999}
{Ludwig}, H.-G., {Freytag}, B., \& {Steffen}, M. 1999, \aap, 346, 111

\bibitem[{{Mahadevan} {et~al.}(2010){Mahadevan}, {Ramsey}, {Wright}, {Endl},
  {Redman}, {Bender}, {Roy}, {Zonak}, {Troupe}, {Engel}, {Sigurdsson},
  {Wolszczan}, \& {Zhao}}]{Mahadevan2010}
{Mahadevan}, S., {Ramsey}, L., {Wright}, J., {et~al.} 2010, in \procspie, Vol.
  7735

\bibitem[{{Maldonado} {et~al.}(2015){Maldonado}, {Affer}, {Micela},
  {Scandariato}, {Damasso}, {Stelzer}, {Barbieri}, {Bedin}, {Biazzo},
  {Bignamini}, {Borsa}, {Claudi}, {Covino}, {Desidera}, {Esposito}, {Gratton},
  {Gonz{\'a}lez Hern{\'a}ndez}, {Lanza}, {Maggio}, {Molinari}, {Pagano},
  {Perger}, {Pillitteri}, {Piotto}, {Poretti}, {Prisinzano}, {Rebolo}, {Ribas},
  {Shkolnik}, {Southworth}, {Sozzetti}, \& {Su{\'a}rez
  Mascare{\~n}o}}]{Maldonado2015}
{Maldonado}, J., {Affer}, L., {Micela}, G., {et~al.} 2015, \aap, 577, A132

\bibitem[{{Mann} {et~al.}(2013{\natexlab{a}}){Mann}, {Brewer}, {Gaidos},
  {L{\'e}pine}, \& {Hilton}}]{Mann2013a}
{Mann}, A.~W., {Brewer}, J.~M., {Gaidos}, E., {L{\'e}pine}, S., \& {Hilton},
  E.~J. 2013{\natexlab{a}}, \aj, 145, 52

\bibitem[{{Mann} {et~al.}(2014){Mann}, {Deacon}, {Gaidos}, {Ansdell}, {Brewer},
  {Liu}, {Magnier}, \& {Aller}}]{Mann2014}
{Mann}, A.~W., {Deacon}, N.~R., {Gaidos}, E., {et~al.} 2014, \aj, 147, 160

\bibitem[{{Mann} {et~al.}(2015){Mann}, {Feiden}, {Gaidos}, {Boyajian}, \& {von
  Braun}}]{Mann2015}
{Mann}, A.~W., {Feiden}, G.~A., {Gaidos}, E., {Boyajian}, T., \& {von Braun},
  K. 2015, \apj, 804, 64

\bibitem[{{Mann} {et~al.}(2013{\natexlab{b}}){Mann}, {Gaidos}, \&
  {Ansdell}}]{Mann2013c}
{Mann}, A.~W., {Gaidos}, E., \& {Ansdell}, M. 2013{\natexlab{b}}, \apj, 779,
  188

\bibitem[{{Mann} {et~al.}(2013{\natexlab{c}}){Mann}, {Gaidos}, {Kraus}, \&
  {Hilton}}]{Mann2013b}
{Mann}, A.~W., {Gaidos}, E., {Kraus}, A., \& {Hilton}, E.~J.
  2013{\natexlab{c}}, \apj, 770, 43

\bibitem[{{McKemmish} {et~al.}(2017){McKemmish}, {Masseron}, {Sheppard},
  {Sandeman}, {Schofield}, {Furtenbacher}, {Cs{\'a}sz{\'a}r}, {Tennyson}, \&
  {Sousa-Silva}}]{McKemmish2017}
{McKemmish}, L.~K., {Masseron}, T., {Sheppard}, S., {et~al.} 2017, \apjs, 228,
  15

\bibitem[{{McLean} {et~al.}(1998){McLean}, {Becklin}, {Bendiksen}, {Brims},
  {Canfield}, {Figer}, {Graham}, {Hare}, {Lacayanga}, {Larkin}, {Larson},
  {Levenson}, {Magnone}, {Teplitz}, \& {Wong}}]{McLean1998}
{McLean}, I.~S., {Becklin}, E.~E., {Bendiksen}, O., {et~al.} 1998, in Infrared
  Astronomical Instrumentation, ed. A.~M. {Fowler}, Vol. 3354, 566--578

\bibitem[{{Mould}(1976)}]{Mould1976}
{Mould}, J.~R. 1976, \apj, 210, 402

\bibitem[{{Mould}(1978)}]{Mould1978}
---. 1978, \apj, 226, 923

\bibitem[{{Neves} {et~al.}(2014){Neves}, {Bonfils}, {Santos}, {Delfosse},
  {Forveille}, {Allard}, \& {Udry}}]{Neves2014}
{Neves}, V., {Bonfils}, X., {Santos}, N.~C., {et~al.} 2014, ArXiv e-prints,
  arXiv:1406.6127

\bibitem[{{Neves} {et~al.}(2012){Neves}, {Bonfils}, {Santos}, {Delfosse},
  {Forveille}, {Allard}, {Nat{\'a}rio}, {Fernandes}, \& {Udry}}]{Neves2012}
---. 2012, \aap, 538, A25

\bibitem[{{Newton} {et~al.}(2014){Newton}, {Charbonneau}, {Irwin},
  {Berta-Thompson}, {Rojas-Ayala}, {Covey}, \& {Lloyd}}]{Newton2014}
{Newton}, E.~R., {Charbonneau}, D., {Irwin}, J., {et~al.} 2014, \aj, 147, 20

\bibitem[{{Newton} {et~al.}(2017){Newton}, {Irwin}, {Charbonneau}, {Berlind},
  {Calkins}, \& {Mink}}]{Newton2017}
{Newton}, E.~R., {Irwin}, J., {Charbonneau}, D., {et~al.} 2017, \apj, 834, 85

\bibitem[{{Newton} {et~al.}(2016){Newton}, {Irwin}, {Charbonneau},
  {Berta-Thompson}, {Dittmann}, \& {West}}]{Newton2016}
---. 2016, \apj, 821, 93

\bibitem[{{Nissen}(2013)}]{Nissen2013}
{Nissen}, P.~E. 2013, \aap, 552, A73

\bibitem[{{Nissen} {et~al.}(2014){Nissen}, {Chen}, {Carigi}, {Schuster}, \&
  {Zhao}}]{Nissen2014}
{Nissen}, P.~E., {Chen}, Y.~Q., {Carigi}, L., {Schuster}, W.~J., \& {Zhao}, G.
  2014, \aap, 568, A25

\bibitem[{{Noyes} {et~al.}(1984){Noyes}, {Hartmann}, {Baliunas}, {Duncan}, \&
  {Vaughan}}]{Noyes1984}
{Noyes}, R.~W., {Hartmann}, L.~W., {Baliunas}, S.~L., {Duncan}, D.~K., \&
  {Vaughan}, A.~H. 1984, \apj, 279, 763

\bibitem[{{Nutzman} \& {Charbonneau}(2008)}]{Nutzman2008}
{Nutzman}, P., \& {Charbonneau}, D. 2008, \pasp, 120, 317

\bibitem[{{{\"O}nehag} {et~al.}(2012){{\"O}nehag}, {Heiter}, {Gustafsson},
  {Piskunov}, {Plez}, \& {Reiners}}]{Onehag2012}
{{\"O}nehag}, A., {Heiter}, U., {Gustafsson}, B., {et~al.} 2012, \aap, 542, A33

\bibitem[{{Passegger} {et~al.}(2016){Passegger}, {Wende-von Berg}, \&
  {Reiners}}]{Passegger2016}
{Passegger}, V.~M., {Wende-von Berg}, S., \& {Reiners}, A. 2016, \aap, 587, A19

\bibitem[{{Petigura} \& {Marcy}(2011)}]{Petigura2011}
{Petigura}, E.~A., \& {Marcy}, G.~W. 2011, \apj, 735, 41

\bibitem[{{Pineda} {et~al.}(2013){Pineda}, {Bottom}, \& {Johnson}}]{Pineda2013}
{Pineda}, J.~S., {Bottom}, M., \& {Johnson}, J.~A. 2013, \apj, 767, 28

\bibitem[{{Piskunov} \& {Valenti}(2017)}]{Piskunov2017}
{Piskunov}, N., \& {Valenti}, J.~A. 2017, \aap, 597, A16

\bibitem[{{Plez}(2012)}]{Plez2012}
{Plez}, B. 2012, {Turbospectrum: Code for spectral synthesis}, Astrophysics
  Source Code Library, ascl:1205.004

\bibitem[{{Quirrenbach} {et~al.}(2010){Quirrenbach}, {Amado}, {Mandel},
  {Caballero}, {Mundt}, {Ribas}, {Reiners}, {Abril}, {Aceituno}, {Afonso},
  {Barrado Y Navascues}, {Bean}, {B{\'e}jar}, {Becerril}, {B{\"o}hm},
  {C{\'a}rdenas}, {Claret}, {Colom{\'e}}, {Costillo}, {Dreizler},
  {Fern{\'a}ndez}, {Francisco}, {Galad{\'{\i}}}, {Garrido}, {Gonz{\'a}lez
  Hern{\'a}ndez}, {Gu{\`a}rdia}, {Guenther}, {Guti{\'e}rrez-Soto}, {Joergens},
  {Hatzes}, {Helmling}, {Henning}, {Herrero}, {K{\"u}rster}, {Laun}, {Lenzen},
  {Mall}, {Martin}, {Mart{\'{\i}}n-Ruiz}, {Mirabet}, {Montes}, {Morales},
  {Morales Mu{\~n}oz}, {Moya}, {Naranjo}, {Rabaza}, {Ram{\'o}n}, {Rebolo},
  {Reffert}, {Rodler}, {Rodr{\'{\i}}guez}, {Rodr{\'{\i}}guez Trinidad},
  {Rohloff}, {S{\'a}nchez Carrasco}, {Schmidt}, {Seifert}, {Setiawan},
  {Solano}, {Stahl}, {Storz}, {Su{\'a}rez}, {Thiele}, {Wagner}, {Wiedemann},
  {Zapatero Osorio}, {Del Burgo}, {S{\'a}nchez-Blanco}, \&
  {Xu}}]{Quirrenbach2010}
{Quirrenbach}, A., {Amado}, P.~J., {Mandel}, H., {et~al.} 2010, in \procspie,
  Vol. 7735

\bibitem[{{Rajpurohit} {et~al.}(2017){Rajpurohit}, {Allard}, {Teixeira},
  {Homeier}, {Rajpurohit}, \& {Mousis}}]{2017arXiv170806211R}
{Rajpurohit}, A.~S., {Allard}, F., {Teixeira}, G.~D.~C., {et~al.} 2017, ArXiv
  e-prints, arXiv:1708.06211

\bibitem[{{Rajpurohit} {et~al.}(2013){Rajpurohit}, {Reyl{\'e}}, {Allard},
  {Homeier}, {Schultheis}, {Bessell}, \& {Robin}}]{Rajpurohit2013}
{Rajpurohit}, A.~S., {Reyl{\'e}}, C., {Allard}, F., {et~al.} 2013, \aap, 556,
  A15

\bibitem[{{Reiners} \& {Basri}(2007)}]{Reiners2007}
{Reiners}, A., \& {Basri}, G. 2007, \apj, 656, 1121

\bibitem[{{Ricker}(2014)}]{Ricker2014}
{Ricker}, G.~R. 2014, Journal of the American Association of Variable Star
  Observers (JAAVSO), 42, 234

\bibitem[{{Rojas-Ayala} {et~al.}(2010){Rojas-Ayala}, {Covey}, {Muirhead}, \&
  {Lloyd}}]{Rojas2010}
{Rojas-Ayala}, B., {Covey}, K.~R., {Muirhead}, P.~S., \& {Lloyd}, J.~P. 2010,
  \apjl, 720, L113

\bibitem[{{Rojas-Ayala} {et~al.}(2012){Rojas-Ayala}, {Covey}, {Muirhead}, \&
  {Lloyd}}]{Rojas2012}
---. 2012, \apj, 748, 93

\bibitem[{{Santos} {et~al.}(2001){Santos}, {Israelian}, \&
  {Mayor}}]{Santos2001}
{Santos}, N.~C., {Israelian}, G., \& {Mayor}, M. 2001, \aap, 373, 1019

\bibitem[{{Savitzky} \& {Golay}(1964)}]{Savitzky1964}
{Savitzky}, A., \& {Golay}, M.~J.~E. 1964, Analytical Chemistry, 36, 1627

\bibitem[{{Schlaufman} \& {Laughlin}(2010)}]{Schlaufman2010}
{Schlaufman}, K.~C., \& {Laughlin}, G. 2010, \aap, 519, A105+

\bibitem[{{Shulyak} {et~al.}(2014){Shulyak}, {Reiners}, {Seemann}, {Kochukhov},
  \& {Piskunov}}]{Shulyak2014}
{Shulyak}, D., {Reiners}, A., {Seemann}, U., {Kochukhov}, O., \& {Piskunov}, N.
  2014, \aap, 563, A35

\bibitem[{{Sneden}(1973)}]{Sneden1973}
{Sneden}, C.~A. 1973, PhD thesis, THE UNIVERSITY OF TEXAS AT AUSTIN.

\bibitem[{{Sousa} {et~al.}(2011){Sousa}, {Santos}, {Israelian}, {Mayor}, \&
  {Udry}}]{Sousa2011}
{Sousa}, S.~G., {Santos}, N.~C., {Israelian}, G., {Mayor}, M., \& {Udry}, S.
  2011, \aap, 533, A141

\bibitem[{{Souto} {et~al.}(2017){Souto}, {Cunha},
  {Garc{\'{\i}}a-Hern{\'a}ndez}, {Zamora}, {Allende Prieto}, {Smith},
  {Mahadevan}, {Blake}, {Johnson}, {J{\"o}nsson}, {Pinsonneault}, {Holtzman},
  {Majewski}, {Shetrone}, {Teske}, {Nidever}, {Schiavon}, {Sobeck},
  {Garc{\'{\i}}a P{\'e}rez}, {G{\'o}mez Maqueo Chew}, \& {Stassun}}]{Souto2017}
{Souto}, D., {Cunha}, K., {Garc{\'{\i}}a-Hern{\'a}ndez}, D.~A., {et~al.} 2017,
  \apj, 835, 239

\bibitem[{{Tennyson} {et~al.}(2016){Tennyson}, {Lodi}, {McKemmish}, \&
  {Yurchenko}}]{Tennyson2016}
{Tennyson}, J., {Lodi}, L., {McKemmish}, L.~K., \& {Yurchenko}, S.~N. 2016,
  Journal of Physics B Atomic Molecular Physics, 49, 102001

\bibitem[{{Terrien} {et~al.}(2012){Terrien}, {Mahadevan}, {Bender},
  {Deshpande}, {Ramsey}, \& {Bochanski}}]{Terrien2012}
{Terrien}, R.~C., {Mahadevan}, S., {Bender}, C.~F., {et~al.} 2012, \apjl, 747,
  L38

\bibitem[{{Terrien} {et~al.}(2015){Terrien}, {Mahadevan}, {Deshpande}, \&
  {Bender}}]{Terrien2015b}
{Terrien}, R.~C., {Mahadevan}, S., {Deshpande}, R., \& {Bender}, C.~F. 2015,
  \apjs, 220, 16

\bibitem[{{Teske} {et~al.}(2014){Teske}, {Cunha}, {Smith}, {Schuler}, \&
  {Griffith}}]{Teske2014}
{Teske}, J.~K., {Cunha}, K., {Smith}, V.~V., {Schuler}, S.~C., \& {Griffith},
  C.~A. 2014, \apj, 788, 39

\bibitem[{{Tsuji} \& {Nakajima}(2014)}]{Tsuji2014}
{Tsuji}, T., \& {Nakajima}, T. 2014, \pasj, 66, 98

\bibitem[{{Tsuji} {et~al.}(2015){Tsuji}, {Nakajima}, \& {Takeda}}]{Tsuji2015}
{Tsuji}, T., {Nakajima}, T., \& {Takeda}, Y. 2015, \pasj, 67, 26

\bibitem[{{Valenti} \& {Piskunov}(1996)}]{Valenti1996}
{Valenti}, J.~A., \& {Piskunov}, N. 1996, \aaps, 118, 595

\bibitem[{{Valenti} {et~al.}(1998){Valenti}, {Piskunov}, \&
  {Johns-Krull}}]{Valenti1998}
{Valenti}, J.~A., {Piskunov}, N., \& {Johns-Krull}, C.~M. 1998, \apj, 498, 851

\bibitem[{{Veyette} {et~al.}(2016){Veyette}, {Muirhead}, {Mann}, \&
  {Allard}}]{Veyette2016b}
{Veyette}, M.~J., {Muirhead}, P.~S., {Mann}, A.~W., \& {Allard}, F. 2016, \apj,
  828, 95

\bibitem[{{Wang} \& {Fischer}(2015)}]{Wang2015}
{Wang}, J., \& {Fischer}, D.~A. 2015, \aj, 149, 14

\bibitem[{{Woolf} \& {Wallerstein}(2005)}]{Woolf2005}
{Woolf}, V.~M., \& {Wallerstein}, G. 2005, \mnras, 356, 963

\bibitem[{{Zhu} {et~al.}(2016){Zhu}, {Wang}, \& {Huang}}]{Zhu2016}
{Zhu}, W., {Wang}, J., \& {Huang}, C. 2016, \apj, 832, 196

\end{thebibliography}

\clearpage

\begin{longrotatetable}
\begin{deluxetable*}{l*{18}{c}}
\tabletypesize{\scriptsize}
\tablewidth{0pt} 
\tablecaption{Measured indices and EWs of calibration sample \label{sample_ews}}
\tablehead{
\colhead{Name} & \colhead{1\tablenotemark{a}} & \colhead{2} & \colhead{3} & \colhead{4} & \colhead{5} & \colhead{6} & \colhead{7} & \colhead{8} & \colhead{9} & \colhead{10} & \colhead{11} & \colhead{12} & \colhead{13} & \colhead{14} & \colhead{15} & \colhead{16} & \colhead{17} & \colhead{18}}
\startdata
PM I02441+4913W & 1.040 & 0.186 & 0.109 & 0.093 & 0.073 & 0.098 & 0.061 & 0.050 & 0.105 & 0.092 & 0.094 & 0.262 & 0.259 & 0.276 & 0.086 & 0.129 & 0.181 & 0.126 \\
PM I02555+2652 & 1.055 & 0.234 & 0.096 & 0.089 & 0.063 & 0.097 & 0.069 & 0.041 & 0.126 & 0.116 & 0.093 & 0.273 & 0.263 & 0.256 & 0.086 & 0.138 & 0.194 & 0.148 \\
Gl 166C & 1.064 & 0.241 & 0.064 & 0.099 & 0.056 & 0.073 & 0.054 & 0.030 & 0.149 & 0.116 & 0.085 & 0.363 & 0.347 & 0.322 & 0.094 & 0.157 & 0.246 & 0.176 \\
PM I05415+5329 & 1.038 & 0.216 & 0.143 & 0.108 & 0.099 & 0.114 & 0.084 & 0.067 & 0.119 & 0.107 & 0.102 & 0.278 & 0.274 & 0.327 & 0.099 & 0.133 & 0.196 & 0.143 \\
PM I06461+3233 & 1.035 & 0.163 & 0.083 & 0.090 & 0.047 & 0.083 & 0.053 & 0.037 & 0.103 & 0.090 & 0.074 & 0.261 & 0.245 & 0.262 & 0.068 & 0.112 & 0.165 & 0.120 \\
PM I07191+6644N & 1.033 & 0.239 & 0.201 & 0.119 & 0.120 & 0.143 & 0.100 & 0.075 & 0.107 & 0.102 & 0.115 & 0.270 & 0.261 & 0.404 & 0.103 & 0.132 & 0.205 & 0.132 \\
PM I08143+6304 & 1.035 & 0.192 & 0.107 & 0.086 & 0.076 & 0.085 & 0.055 & 0.056 & 0.105 & 0.092 & 0.085 & 0.277 & 0.261 & 0.279 & 0.082 & 0.125 & 0.187 & 0.135 \\
PM I08526+2818 & 1.062 & 0.241 & 0.093 & 0.085 & 0.063 & 0.100 & 0.071 & 0.038 & 0.142 & 0.117 & 0.079 & 0.286 & 0.271 & 0.261 & 0.094 & 0.141 & 0.210 & 0.149 \\
PM I09151+2321S & 1.035 & 0.231 & 0.186 & 0.122 & 0.103 & 0.138 & 0.096 & 0.079 & 0.115 & 0.109 & 0.119 & 0.280 & 0.256 & 0.402 & 0.098 & 0.139 & 0.204 & 0.144 \\
PM I09573+5018 & 1.028 & 0.167 & 0.118 & 0.090 & 0.070 & 0.093 & 0.059 & 0.055 & 0.092 & 0.086 & 0.092 & 0.248 & 0.238 & 0.310 & 0.077 & 0.109 & 0.172 & 0.117 \\
PM I11218+1811 & 1.032 & 0.259 & 0.193 & 0.119 & 0.103 & 0.133 & 0.096 & 0.083 & 0.124 & 0.121 & 0.119 & 0.290 & 0.264 & 0.409 & 0.111 & 0.148 & 0.217 & 0.153 \\
LSPM J1140+0930E & 1.040 & 0.188 & 0.099 & 0.096 & 0.085 & 0.087 & 0.051 & 0.048 & 0.115 & 0.094 & 0.078 & 0.272 & 0.277 & 0.285 & 0.090 & 0.127 & 0.183 & 0.131 \\
PM I13113+0936 & 1.022 & 0.144 & 0.107 & 0.071 & 0.060 & 0.076 & 0.051 & 0.052 & 0.074 & 0.073 & 0.084 & 0.216 & 0.214 & 0.310 & 0.073 & 0.100 & 0.149 & 0.107 \\
PM I13168+1700 & 1.031 & 0.168 & 0.097 & 0.088 & 0.077 & 0.090 & 0.059 & 0.050 & 0.108 & 0.094 & 0.085 & 0.255 & 0.245 & 0.288 & 0.078 & 0.110 & 0.172 & 0.117 \\
PM I13314-0759W & 1.026 & 0.180 & 0.137 & 0.093 & 0.086 & 0.101 & 0.068 & 0.064 & 0.101 & 0.088 & 0.092 & 0.259 & 0.248 & 0.321 & 0.085 & 0.113 & 0.175 & 0.131 \\
LSPM J1404+0157 & 1.043 & 0.201 & 0.118 & 0.092 & 0.090 & 0.097 & 0.062 & 0.065 & 0.113 & 0.101 & 0.094 & 0.277 & 0.273 & 0.296 & 0.094 & 0.143 & 0.190 & 0.155 \\
PM I14182+1244W & 1.021 & 0.123 & 0.055 & 0.059 & 0.046 & 0.054 & 0.041 & 0.034 & 0.072 & 0.070 & 0.067 & 0.233 & 0.225 & 0.248 & 0.059 & 0.100 & 0.158 & 0.095 \\
PM I14206-2323N & 1.033 & 0.233 & 0.172 & 0.114 & 0.100 & 0.129 & 0.084 & 0.073 & 0.109 & 0.097 & 0.124 & 0.265 & 0.264 & 0.379 & 0.100 & 0.133 & 0.196 & 0.140 \\
PM I15118+3933 & 1.047 & 0.213 & 0.086 & 0.081 & 0.072 & 0.080 & 0.056 & 0.046 & 0.110 & 0.101 & 0.080 & 0.291 & 0.275 & 0.258 & 0.086 & 0.128 & 0.193 & 0.140 \\
PM I15164+1647W & 1.021 & 0.176 & 0.149 & 0.095 & 0.078 & 0.103 & 0.059 & 0.065 & 0.086 & 0.073 & 0.082 & 0.213 & 0.227 & 0.340 & 0.070 & 0.103 & 0.163 & 0.106 \\
PM I15204+0011 & 1.025 & 0.154 & 0.110 & 0.091 & 0.049 & 0.074 & 0.053 & 0.051 & 0.089 & 0.073 & 0.078 & 0.242 & 0.216 & 0.309 & 0.066 & 0.097 & 0.154 & 0.103 \\
PM I16072-1422 & 1.032 & 0.238 & 0.195 & 0.123 & 0.111 & 0.128 & 0.098 & 0.066 & 0.099 & 0.098 & 0.112 & 0.245 & 0.257 & 0.400 & 0.087 & 0.122 & 0.188 & 0.161 \\
PM I16139+3346 & 1.048 & 0.220 & 0.099 & 0.088 & 0.077 & 0.091 & 0.063 & 0.049 & 0.111 & 0.096 & 0.085 & 0.293 & 0.280 & 0.273 & 0.085 & 0.135 & 0.197 & 0.147 \\
PM I16148+6038 & 1.055 & 0.246 & 0.069 & 0.081 & 0.064 & 0.090 & 0.050 & 0.037 & 0.134 & 0.102 & 0.070 & 0.292 & 0.316 & 0.271 & 0.080 & 0.140 & 0.210 & 0.131 \\
PM I17176+5224 & 1.063 & 0.257 & 0.074 & 0.097 & 0.067 & 0.091 & 0.068 & 0.038 & 0.147 & 0.119 & 0.079 & 0.319 & 0.311 & 0.290 & 0.091 & 0.146 & 0.217 & 0.150 \\
LSPM J1742+1643 & 1.035 & 0.203 & 0.109 & 0.090 & 0.075 & 0.088 & 0.065 & 0.057 & 0.113 & 0.103 & 0.072 & 0.278 & 0.256 & 0.304 & 0.083 & 0.119 & 0.183 & 0.173 \\
PM I17464+2743W & 1.051 & 0.221 & 0.106 & 0.091 & 0.081 & 0.095 & 0.065 & 0.058 & 0.118 & 0.106 & 0.095 & 0.293 & 0.273 & 0.269 & 0.093 & 0.145 & 0.205 & 0.155 \\
PM I18006+6832 & 1.031 & 0.231 & 0.183 & 0.113 & 0.108 & 0.131 & 0.089 & 0.078 & 0.105 & 0.092 & 0.100 & 0.238 & 0.234 & 0.377 & 0.089 & 0.116 & 0.180 & 0.105 \\
PM I18007+2933 & 1.040 & 0.198 & 0.095 & 0.094 & 0.073 & 0.090 & 0.058 & 0.048 & 0.117 & 0.100 & 0.079 & 0.277 & 0.265 & 0.275 & 0.087 & 0.127 & 0.183 & 0.131 \\
\enddata
\tablenotetext{a}{Numbers indicate which feature, corresponding to rows in Table~\ref{features}.}
\tablecomments{Equivalents widths are measured in units of $\AA$.}
\end{deluxetable*}
\end{longrotatetable}

\begin{longrotatetable}
\begin{deluxetable*}{*{21}{c}}
\tabletypesize{\scriptsize}
\tablewidth{0pt} 
\tablecaption{Measured indices and EWs of model grid \label{grid_ews}}
\tablehead{
\colhead{\teff{} [K]} & \colhead{\mh{}} & \colhead{\am{}} & \colhead{1\tablenotemark{a}} & \colhead{2} & \colhead{3} & \colhead{4} & \colhead{5} & \colhead{6} & \colhead{7} & \colhead{8} & \colhead{9} & \colhead{10} & \colhead{11} & \colhead{12} & \colhead{13} & \colhead{14} & \colhead{15} & \colhead{16} & \colhead{17} & \colhead{18}}
\startdata
4200 & +0.50 & +0.40  & 1.035 & 0.240 & 0.157 & 0.115 & 0.128 & 0.128 & 0.088 & 0.086 & 0.127 & 0.126 & 0.145 & 0.313 & 0.274 & 0.444 & 0.112 & 0.139 & 0.248 & 0.145 \\
4200 & +0.50 & +0.30  & 1.037 & 0.245 & 0.163 & 0.117 & 0.134 & 0.134 & 0.091 & 0.089 & 0.124 & 0.122 & 0.141 & 0.303 & 0.261 & 0.428 & 0.106 & 0.135 & 0.239 & 0.140 \\
4200 & +0.50 & +0.20  & 1.039 & 0.252 & 0.169 & 0.120 & 0.140 & 0.140 & 0.094 & 0.092 & 0.121 & 0.118 & 0.137 & 0.294 & 0.249 & 0.412 & 0.100 & 0.132 & 0.230 & 0.136 \\
4200 & +0.50 & +0.10  & 1.041 & 0.259 & 0.174 & 0.123 & 0.146 & 0.146 & 0.097 & 0.095 & 0.118 & 0.114 & 0.133 & 0.285 & 0.237 & 0.396 & 0.095 & 0.129 & 0.222 & 0.131 \\
4200 & +0.50 & +0.00  & 1.043 & 0.266 & 0.180 & 0.127 & 0.153 & 0.152 & 0.100 & 0.098 & 0.115 & 0.110 & 0.129 & 0.276 & 0.225 & 0.380 & 0.089 & 0.126 & 0.215 & 0.128 \\
4200 & +0.50 & -0.10  & 1.045 & 0.275 & 0.186 & 0.130 & 0.159 & 0.158 & 0.103 & 0.101 & 0.113 & 0.107 & 0.125 & 0.268 & 0.214 & 0.364 & 0.084 & 0.123 & 0.208 & 0.124 \\
4200 & +0.25 & +0.40  & 1.033 & 0.208 & 0.138 & 0.105 & 0.109 & 0.114 & 0.075 & 0.076 & 0.116 & 0.117 & 0.133 & 0.296 & 0.262 & 0.418 & 0.102 & 0.131 & 0.231 & 0.134 \\
4200 & +0.25 & +0.30  & 1.035 & 0.214 & 0.143 & 0.108 & 0.115 & 0.119 & 0.078 & 0.079 & 0.113 & 0.113 & 0.129 & 0.287 & 0.250 & 0.403 & 0.097 & 0.128 & 0.223 & 0.129 \\
4200 & +0.25 & +0.20  & 1.037 & 0.220 & 0.149 & 0.111 & 0.121 & 0.125 & 0.080 & 0.082 & 0.110 & 0.109 & 0.124 & 0.277 & 0.237 & 0.387 & 0.091 & 0.124 & 0.215 & 0.125 \\
4200 & +0.25 & +0.10  & 1.039 & 0.226 & 0.154 & 0.114 & 0.126 & 0.130 & 0.083 & 0.085 & 0.107 & 0.105 & 0.119 & 0.267 & 0.225 & 0.370 & 0.086 & 0.121 & 0.207 & 0.120 \\
\enddata
\tablenotetext{a}{Numbers indicate which feature, corresponding to rows in Table~\ref{features}.}
\tablecomments{Equivalents widths are measured in units of $\AA$. \\
Table 5 is published in its entirety in the machine-readable format. A portion is shown here for guidance regarding its form and content.}
\end{deluxetable*}
\end{longrotatetable}

\end{document}